\documentclass[preprint,aps,nofootinbib,preprintnumbers,amsmath,amssymb,11pt]{revtex4}
\usepackage{graphicx}
\usepackage{dcolumn}
\usepackage{bm}
\usepackage{natbib}
\usepackage{empheq}
\usepackage{xcolor}
\usepackage{enumerate}

\newcommand{\p}{\partial}
\newcommand{\half}{\frac{1}{2}}

\newcommand\nn{{\nonumber}}
\newcommand{\beq}{\begin{equation}}
\newcommand{\eq}{\end{equation}}

\def\bea{\begin{eqnarray}}
\def\ea{\end{eqnarray}}
\newcommand{\be}{\begin{equation}}
\newcommand{\ee}{\end{equation}}

\begin{document}

\preprint{CCTP-2016-06}
\preprint{CCQCN-2016-144}

\title{Criteria For Superfluid Instabilities of Geometries \\ with Hyperscaling Violation}

\author{Sera Cremonini $ ^{\,\spadesuit}$}
\affiliation{
\it $ ^\spadesuit$ Department of Physics, Lehigh University,\\
\it Bethlehem, PA, 18018 USA}
\author{Li Li $ ^{\,\clubsuit}$$^\blacklozenge$}
\affiliation{\it $ ^{\,\clubsuit}$ Crete Center for Theoretical Physics,\\
\it $^\blacklozenge$ Crete Center for Quantum Complexity and Nanotechnology, \\
Department of Physics, University of Crete, 71003 Heraklion, Greece}
\date{\today}

\begin{abstract}
We examine the onset of superfluid instabilities for geometries that exhibit hyperscaling violation
and Lifshitz-like scaling at infrared and intermediate energy scales, and approach AdS in the ultraviolet.
In particular, we are interested in the role of a non-trivial coupling between the neutral scalar 
supporting the scaling regime, and the  (charged) complex scalar which condenses.
The analysis focuses exclusively on unstable modes arising
from the hyperscaling-violating portion of the geometry.
Working at zero temperature, we identify simple analytical criteria for the presence of
scalar instabilities, and discuss under which conditions a minimal charge will be needed to trigger a transition.
Finite temperature examples
are constructed numerically for a few illustrative cases.
\end{abstract}

\pacs{Valid PACS appear here}

\maketitle
\thispagestyle{empty}

\newpage
\tableofcontents
\newpage

\section{Introduction and Summary of Results}
\label{Intro}

The recent efforts to use holography to probe strongly coupled quantum systems have led to
new insights into the possible instabilities of a variety of gravitational solutions.
One of the prime examples is that of charged black holes in Anti de Sitter (AdS) space,
which have been understood to be unstable to the formation of scalar hair --
thanks to attempts to realize the spontaneous breaking of an abelian gauge symmetry in gravity \cite{Gubser:2008px},
and develop a holographic description of superconducting\footnote{Strictly speaking, the dual theory
consists of a condensate breaking a global $U(1)$ symmetry,  so the description is of a superfluid
rather than a superconductor. However, considering the limit in which the $U(1)$ symmetry
is "weakly gauged", we can still view the dual theory as a superconductor. In the present paper we will not distinguish between the
two terminologies.} phases \cite{Hartnoll:2008vx,Hartnoll:2008kx}.
For reviews of holographic superconductors we refer the reader to \emph{e.g.}~\cite{Hartnoll:2009sz,Herzog:2009xv,Horowitz:2010gk,Cai:2015cya}.
Other notable examples include the spontaneous breaking of translational invariance and the onset of spatially
modulated instabilities,
which have been identified in a number of geometries (see~\cite{Domokos:2007kt,Nakamura:2009tf,Ooguri:2010xs,Donos:2011bh,Bergman:2011rf}
for some of the early papers) and have potential applications to \emph{e.g.}
QCD and condensed matter systems with striped phases.
We have seen growing interest in constructing gravitational solutions
that exhibit a variety of broken symmetries,
with significant attention recently given to realizing holographic lattices through
the (explicit) breaking of translational invariance
(see \emph{e.g.} \cite{Hartnoll:2012rj,Horowitz:2012ky,
Donos:2012js,Vegh:2013sk,Blake:2013bqa,Donos:2013eha,Andrade:2013gsa,Gouteraux:2014hca,Donos:2014oha}).

In this paper we revisit the question of scalar field instabilities associated with geometries that exhibit
hyperscaling violation $\theta$ and non-relativistic scaling $z$,
with the ultimate goal of reaching a more complete understanding of low temperature superconducting phase transitions in the dual systems.
We will work with gravitational solutions which are hyperscaling violating and Lifshitz-like
 at infrared (IR) and intermediate energies, and asymptote to  AdS in the ultraviolet (UV).
Such geometries are well known to arise in Einstein-Maxwell-dilaton theories, and are
supported by a neutral scalar subject to a rather simple potential.
We require AdS asymptotics to ensure that the dual field theory is conformal at the UV fixed point -- so that the violation of
hyperscaling and relativistic symmetry is generated at lower energies -- and thus can rely on the standard holographic dictionary.
We stress that we are only interested in phase transitions that are triggered in the hyperscaling violating regime itself,
since in full generality they are much less understood than their AdS counterpart.

Charged scalar field condensation on non-relativistic backgrounds that don't respect hyperscaling has been studied in a number of
settings (see \emph{e.g.} \cite{Salvio:2013jia,Fan:2013tga,Lucas:2014sba} but the list is by no means exhaustive), although typically
for specific values of the scaling exponents $z$ and $\theta$ or in somewhat simple models.
Here we will extend these analyses by introducing a non-trivial coupling of the form
$\sim B(\phi) \, |\Psi|^2$ between the neutral scalar $\phi$ that determines the
background and the charged scalar $\Psi$ that condenses.
We will obtain analytical instability criteria -- attempting to be generic, to the extent that it is possible --
and highlight the role of $B(\phi)$ on the onset of the superfluid phase transition.
Since $B(\phi)$ contributes to the effective mass of the charged scalar, it is intuitively clear that it will affect the
condensation process -- enhancing it or impeding it depending on its sign and its radial profile.
Throughout the paper we will adopt the choice $B(\phi) \sim e^{\hat{\tau}\phi}$
in the hyperscaling violating regime, with $\hat\tau$ an arbitrary constant.

To probe the onset of the formation of scalar hair, we are going to focus on the
linearized perturbation of the charged scalar $\Psi$ around the unbroken phase.
To obtain the linearized equation of motion for $\Psi$ it suffices to know the structure of the charged scalar couplings up to quadratic order --
such leading terms are enough to compute the temperature at which the unbroken phase becomes unstable to scalar hair.
One should keep in mind, however, that the nonlinear details of the couplings could affect the order of the phase transition
and the thermodynamics, as has been stressed in~\cite{Kiritsis:2015hoa}.

Our instability analysis will be done in two complementary ways.
After setting up the model and the background in Sections \ref{Setup} and \ref{BackgroundSection},
we will inspect the behavior of the effective mass $M_{eff}^2$ of the charged scalar in Section \ref{masssection}, and
in particular, the conditions under which it becomes sufficiently negative.
In Section \ref{SchrodingerSection} we will then recast the linearized perturbation of the charged scalar in Schr\"{o}dinger form,
and perform a more detailed instability analysis by examining whether the effective Schr\"{o}dinger potential $V_{Schr}$
is sufficiently negative to support bound states (for studies of instabilities
in terms of an effective Schr\"{o}dinger potential see \emph{e.g.} \cite{Kodama:2003kk,Hartnoll:2008kx,Keeler:2013msa}).
To complement the intuition developed from examining $M_{eff}^2$ and $V_{Schr}$, one should also analyze the structure of IR
perturbations of the charged scalar, to ensure that they can indeed support a scalar condensate.
As we will see, this can rule out regions of parameter space for which $M_{eff}^2$ and $V_{Schr}$ may be ambiguous.
For simplicity, our analytical arguments are developed working at zero temperature, and are meant to serve as guidance for a more
detailed finite temperature analysis. Still, we believe that they capture all the essential physics of their low temperature counterpart,
as we confirm in our numerical section \ref{NumericsSection}, in a few illustrative cases.
We leave a more thorough finite temperature analysis to future work.

We will find many similarities with the standard holographic superconductor setup, but also some crucial differences.
As in \cite{Gubser:2008px,Hartnoll:2008vx,Hartnoll:2008kx}, two distinct mechanisms can lead to the condensation of a scalar in these background
geometries.
The gauge field contribution to the effective mass $M^2_{eff}$ of $\Psi$ is always negative and
can become large enough to make it energetically favorable for the system to undergo a superfluid phase transition.
Similarly, a negative coupling $B(\phi)$ can drive $M_{eff}^2$ to become appreciably negative, thus
facilitating the transition.
Since the latter process can happen even at zero charge, it allows neutral scalars to condense -- and it is
of course the analog of violating BF bounds in AdS.

What is novel in the models we consider here is the rich behavior associated with the possible
profiles of the coupling $B(\phi)$, and its effect on the interplay between the two instability mechanisms.
In particular, the condensation process is highly sensitive to the specific way in which $B(\phi)$ scales
as compared to the $\{z,\theta\}$ background geometry
-- qualitatively new behavior will be seen when the effective mass term $B(\phi) \, \psi^2$
does \emph{not} respect the scaling of the charged scalar kinetic term (here
 $\psi$ denotes the modulus of the complex scalar $\Psi$).
We should note that the role of a coupling  $\sim B(\phi) \psi^2$ in hyperscaling violating backgrounds
was already discussed by \cite{Lucas:2014sba}, although in a slightly different context.
Choosing the coupling so that $B(\phi) \psi^2 $ scales as $\sim (\p\psi)^2$,
the authors noted the presence of a minimal charge needed to form a condensate,
and raised the question of whether it could be a universal feature.
Here we will address this point working with general classes of $\{z,\theta\}$ geometries and couplings
$B(\phi) \sim e^{\hat{\tau}\phi}$, and show that this is not generally the case -- there is a somewhat large parameter space where neutral
scalars can condense.
We will also identify the cases in which we expect to see a minimal charge.
As we will see, the existence of the latter will be sensitive to the detailed behavior of $B(\phi)$.
Again, we find some crucial differences with the standard holographic superconductor setup\footnote{However,
see \cite{Kiritsis:2015hoa,Iqbal:2010eh} for additional ways to modify the effective mass of a charged scalar in the IR $AdS_2$ region.},
that can be traced to the non-trivial scaling properties of the coupling $B$ and the background itself.


\subsection{Summary of Results}
We work with the Lagrangian given in (\ref{Lag}),
so that the dual field theory has $d$ spatial dimensions.
To respect the scaling of the
potential $V(\phi) \propto e^{-\beta\phi}$ and gauge kinetic function $Z(\phi)\propto e^{\, \alpha\phi}$
of the hyperscaling violating background,
we have taken the coupling between the two scalars to be of the form
$B(\phi) \sim e^{\hat{\tau}\phi}$ or, in terms of the holographic radial coordinate $r$,
\beq
B(r) = B_0 \, r^\tau \, ,
\eq
with $\tau$ an arbitrary constant.

Our analytical estimates for the onset of scalar field instabilities
are extracted first in Section \ref{masssection} by inspecting the
effective mass for the charged scalar
\beq
M_{eff}^2(r) = \tilde{L}^2 \left[ B_0 \, r^{\tau -2(m-1)} - Q^2 r^{2dn}\right] + \Bigl(m+\half d n\Bigr) \Bigl(m+\half d n-1\Bigr) \, ,
\eq
and then in Section \ref{SchrodingerSection} by examining when the effective Schr\"{o}dinger potential
\beq
V_{Schr}(r) =r^{2(2m-1)}\left[ B_0 \, r^{\tau-2(m-1)}-Q^2 r^{2 d n} +
\frac{1}{4\tilde{L}^2} \, dn \Bigl(dn+4m-2\Bigr) \right]
\eq
develops negative regions which can support the existence of bound states.
Here $Q$ is proportional to the charge of $\Psi$, $\tilde{L}$ is a length scale defined in the main text
and the parameters $\{m,n\}$ are
\beq
m = \frac{z d -\theta}{z d - 2\theta} \, , \qquad n = \frac{d -\theta}{z d - 2\theta} \, .
\eq
With our choice of coordinates the IR is located at $r=0$, while $r=r_{tr}$ will denote the transition scale between the
non-relativistic, hyperscaling violating solution and the UV AdS region.

The possible sources of instability are now apparent.
Superfluid phase transitions are generically triggered by a sufficiently large charge term $\propto Q^2$, driving $M_{eff}$ imaginary and $V_{Schr}$ negative.
A negative and suitably large contribution from the coupling $\propto B_0$ will have the same effect, and is responsible for the
formation of a condensate even when $Q=0$.
Moreover, the interplay between the two terms can lead to interesting behaviors, depending on how $\tau$ compares to the exponents $m$ and $n$.
While these expressions were obtained at zero temperature, they are expected to capture the key aspects of the finite temperature
behavior. This is shown for a few illustrative cases in Section \ref{NumericsSection}.\\

\noindent
The main features that have emerged from this analysis are the following:
\begin{itemize}
\item
In these hyperscaling violating backgrounds the gauge field term $\sim Q^2 r^{2dn}$ always decreases towards to IR (as $r\rightarrow 0$),
since $n>0$, as discussed in the main text.
\item
The competition between the contributions coming from the $U(1)$ gauge field and the real neutral scalar is very sensitive
to the way in which $B(\phi)$ scales compared to the background, in particular to whether $\tau$ is larger or smaller than $2(m-1)$.
\item
Simplifications occur for the \emph{scaling} choice $\tau = 2(m-1)$, which corresponds to the coupling $B(\phi) \psi^2$
scaling in the same way as the kinetic term $(\p \psi)^2$.
In this case the only radial dependence of $M_{eff}^2$ comes from the charge term, and in the deep IR one obtains generalized BF bounds
analogous to those in AdS.
\item
In the scaling case $\tau = 2(m-1)$ neutral scalars will condense when $B_0$ is sufficiently negative.
There will otherwise be a minimal charge $Q_{min}$ needed to trigger the condensation, as in the standard AdS case.
However, here bound states are supported near the transition region $r \sim r_{tr}$ to AdS, and instabilities are therefore associated with the
``effective UV'' of the $\{z,\theta\}$ geometry, and not with its IR.
\item
When the scaling $\tau$ is arbitrary the behavior is more complex:
\begin{enumerate}[(i)]
\item
For $B_0<0$ and $\tau<2(m-1)$ the coupling makes $V_{Schr}$ and $M^2_{eff}$ more and more negative as the IR is approached.
Thus, neutral scalars will condense generically, without having to tune the size of $B_0$, unlike in the standard AdS case.
The instability is now associated with the IR of the geometry, and there is no minimal charge.
\item
In all other cases a minimal charge seems to be needed to trigger the phase transition.
A particularly interesting case corresponds to $B_0<0$ and $\tau > 2(m-1)$. Here $Q_{min}$ exists independently of how large
 $|B_0|$ is tuned to be, unlike in the standard AdS story.
\end{enumerate}
\item
The choice $\tau - 2(m-1) = 2dn$ is also special, since the coupling and charge contributions to $M_{eff}^2$ and $V_{Schr}$
scale in the same way, $\propto r^{2dn}\left[B_0 - Q^2 \right]$:
\begin{enumerate}[(i)]
\item
For $B_0 > Q^2$ there will never be a phase transition  
triggered in the IR hyperscaling violating region, no matter how large the charge is.
\item
For $B_0 < Q^2$ we expect to have a condensate, as long as the effective mass can become negative enough near $r_{tr}$, where $r^{2dn}$
attains its largest value. Thus, one can trigger a transition by varying $B_0$ across the critical value $Q^2$.
However, there will always be a minimal charge, no matter how negative $B_0$ is.
\end{enumerate}
\item
The transition scale $r_{tr}$ between the hyperscaling violating geometry and the AdS region plays a crucial role in controlling the onset of
the instability and the value of the minimal charge.
\end{itemize}

\section{Setup}
\label{Setup}

We want to examine $D=d+2$ dimensional Einstein-Maxwell-dilaton theories coupled to a complex scalar field $\Psi$,
\beq
\label{Lag}
\mathcal{L}_{d+2} = R  - \half (\p\phi)^2 - \frac{1}{4} Z(\phi) F_{\mu\nu}F^{\mu\nu} - V(\phi)  -
C(\phi)\left( |D\Psi|^2 + B(\phi) |\Psi|^2 \right) \, ,
\eq
which is charged under the $U(1)$ field $A_\mu$, so that $D_\mu \Psi = (\p_\mu +i q A_\mu)\Psi$.
For now we allow for two arbitrary couplings $C(\phi)$ and $B(\phi)$
between the neutral and the charged scalars. The former results in a non-canonical kinetic term
for $\Psi$ and will be set to one shortly, the latter acts as an effective mass for $\Psi$ and will be the focus of our discussion.

Einstein's equations for~\eqref{Lag} are given by
\bea
&& R_{\mu\nu} + \half \, Z(\phi) \, F_{\mu\rho} F^\rho_{\; \;\nu} - \half \p_\mu\phi \, \p_\nu\phi
- \frac{1}{2}C(\phi) \, \left[D_\mu \Psi (D_\nu \Psi)^\ast + D_\nu \Psi (D_\mu \Psi)^\ast\right] \nn\\
&& \quad +  \half g_{\mu\nu} \left[ \half (\p\phi)^2 +V(\phi)-R + \frac{Z}{4} F^2 + C(\phi)
\left(  |D\Psi|^2 + B(\phi) |\Psi|^2 \right) \right] = 0 \, ,
\ea
where, writing the charged scalar as $\Psi = \psi e^{i\theta}$, we have
\beq
|D\Psi|^2 = \left[ (\p\psi)^2 +\psi^2(\p\theta+q A)^2  \right] \, .
\eq
The gauge field equation of motion is
\bea
\frac{1}{\sqrt{-g}} \p_\mu \left( \sqrt{-g} Z F^{\mu\nu} \right) &=& 2 \, C\, q^2 A^\nu |\Psi|^2
+ i \, q \, C \left[ \Psi \p^\nu \Psi^* - \Psi^* \p^\nu \Psi \right] \nn \\
&=& 2 \,C \, q^2 A^\nu \psi^2 + 2 \, C \, q \, \psi^2 \p^\nu \theta \, ,
\ea
while the neutral scalar obeys
\bea
\Box\phi &=& \frac{\p V}{\p \phi} + \frac{1}{4} \frac{\p Z}{\p \phi} F^2 + \frac{\p C}{\p \phi} |D\Psi|^2 +
\frac{\p (C \, B)}{\p \phi} \, |\Psi|^2 \, .
\ea
Finally, the real $\psi$ and imaginary $\theta$ parts of the charged scalar satisfy
\bea
&& \frac{1}{\sqrt{-g} \, C}\p_\mu \left( \sqrt{-g} \, C \p^\mu \psi \right) =
\left[ (\p\theta+q A)^2  + B\right] \psi \, , \\
\label{thetaEOM}
&& \p_\mu \left[ \sqrt{-g} \,C\, \psi^2 \left( \p^\mu \theta + q A^\mu \right) \right] = 0 \, .
\ea
We take the phase of the charged scalar to vanish, $\theta = 0$.
This solves the equation of motion (\ref{thetaEOM}) when
the gauge field is purely electric, $A = A_t (r) dt$ and no fields depend explicitly on time.
The charged scalar equation of motion then becomes
\beq
\label{chargedscalarEOM}
\frac{1}{\sqrt{-g} \; C (\phi)}\p_\mu \left( \sqrt{-g} \; C(\phi) \, \p^\mu \psi \right) =
\left[ q^2 A_\mu A^\mu  + B(\phi) \right] \psi  \, .
\eq
While the non-canonicality function $C(\phi)$ could contribute to the instabilities in an interesting way\footnote{Superconducting/Superfluid
instabilities in a theory with non-canonical couplings has been considered recently on top of soft wall backgrounds in~\cite{Cubrovic:2016uuy}.} --
it clearly affects the scaling behavior of the charged
scalar, and hence the scaling dimension of the dual operator --
here for simplicity we will neglect it  and set $C=1$, focusing instead on the role of the $B(\phi)$ coupling.
We then see that (\ref{chargedscalarEOM}) becomes
\beq
\Box \psi =  m_{eff}^2 \, \psi \, ,
\eq
with the effective mass given by
\beq
\label{meff}
m_{eff}^2 = q^2 A_\mu A^\mu  + B(\phi) =
- q^2 A_t^2 |g^{tt}| + B(\phi) \, .
\eq
As in the case of the standard holographic superconductor \cite{Gubser:2008px,Hartnoll:2008vx,Hartnoll:2008kx},
the condensation of the charged scalar field will depend on the interplay between the two contributions to its effective mass,
one coming from the coupling $B(\phi)$ and the other from the charge.
However, we will see that the additional dependence on the neutral scalar 
-- and in particular, the fact that the profile of $B(\phi)$ will depend on the holographic radial coordinate and can be chosen to scale in
different ways --
will lead to some interesting differences.

\section{Background geometry}
\label{BackgroundSection}

The instability we are interested in is associated with the formation of charged scalar hair
around the normal unbroken black brane background in which 
$\psi$ is zero.
In the vicinity of the transition point at which scalar hair begins to develop,
the value of $\psi$ should be very small, and backreaction negligible.
As a result we can treat the charged scalar as a perturbation on top of the background solution
which interpolates between asymptotic AdS and a Lifshitz-like, hyperscaling violating region that extends into the IR.
More precisely, the latter geometry extends over the range $r_{IR} < r < r_{tr}$, with AdS describing the
remaining $r_{tr} < r < r_{UV}$ portion of the spacetime.
Thus, $r_{tr}$ denotes the transition scale between the two regimes, while $r_{IR}$ and $r_{UV}$ correspond to
the IR and UV endpoints of RG flow.
In what follows we set the charged scalar field to zero, and focus entirely on the background geometry.

\subsection{The hyperscaling violating background solution}

We begin by discussing the non-relativistic $\{z,\theta\}$ scaling solutions that range over the infrared and intermediate part of the geometry.
It is well known that such solutions can be generated in the class of models (\ref{Lag}) by taking the scalar potential and gauge kinetic function
to be simple exponentials,
\beq
\label{VandZ}
Z(\phi) = Z_0 e^{\alpha\phi} \, , \quad \quad \quad  V (\phi) = -V_0 e^{-\beta\phi} \, ,
\eq
where $V_0$ and $Z_0$ are arbitrary positive constants.
Black brane solutions are then given by \cite{Charmousis:2010zz,Iizuka:2011hg,Gouteraux:2011ce,Huijse:2011ef}
\begin{equation}\label{hyperscalz}
\begin{split}
& ds^2_{\text{ h.v.}}=\rho^{\frac{2\theta}{d}}\left(-f(\rho)\frac{dt^2}{\rho^{2 z}}
+\frac{L^2}{\rho^2}\frac{d\rho^2}{f(\rho)} + \frac{d\vec{x}^{2}}{\rho^2} \right) \, ,\\
& f(\rho)=1-\left(\frac{\rho}{\rho_h}\right)^{d+z-\theta} \, , \qquad L^2=\frac{(d-1+z-\theta)(d+z-\theta)}{V_0}\, , \\
& \alpha=\frac{2d}{\kappa}-\frac{2(d-1)\theta}{d \kappa},\qquad \beta=\frac{2\theta}{d\kappa} \, ,
\qquad \kappa^2 = \frac{2(\theta-d)(\theta-dz+d)}{d} \, ,
\end{split}
\end{equation}
and are supported by the following scalar and gauge field profiles
\beq
\phi = \kappa \ln (\rho) \, , \qquad A=a_0 \, \rho^{\theta-z-d}f(\rho) \, dt \, , \qquad a_0 =\sqrt{\frac{2(z-1)}{Z_0 (d+z-\theta)}} \, .
\eq
In the extreme limit the metric reduces to
\beq
\label{hsvzeroz}
ds^2_{\text{ h.v.}}=\rho^{\frac{2\theta}{d}}\left(-\frac{dt^2}{\rho^{2 z}} + L^2 \frac{d\rho^2}{\rho^2} +
\frac{d\vec{x}^{2}}{\rho^2} \right) \, ,
\eq
and is known to suffer generically from curvature and null singularities.
Moreover, the logarithmically running scalar $\phi \sim \ln\rho$
tends to drive the bulk gravitational theory to strong or weak coupling, depending on whether the gauge field
is chosen to describe a magnetic or an electric field.
Although a possible resolution comes from turning on a temperature,
the presence of instabilities in these systems generically indicates that there may be additional ground states.
Possible IR completions of these scaling geometries
have been discussed \emph{e.g.} in \cite{Harrison:2012vy,Bhattacharya:2012zu,Kundu:2012jn,Cremonini:2012ir,
Iizuka:2013ag,Knodel:2013fua,Cremonini:2013epa,Barisch-Dick:2013xga,O'Keeffe:2013nha,Bhattacharya:2014dea}.

There are some disadvantages to using the $\rho$ radial coordinate adopted above.
For example\footnote{In order to have an unambiguous IR one must require
$\left(\frac{2\theta}{d}-2z\right)\left(\frac{2\theta}{d}-2\right)>0$,
which simply ensures that the $(t, \vec{x})$ components of the metric scale in the same way with $\rho$.},
whether the IR is located at $\rho=0$ or $\rho \rightarrow \infty$ depends on the values of $\{z,\theta\}$.
Also, in these coordinates in order to recover the standard $AdS_2 \times \mathbb{R}^d$ extremal solution to Einstein-Maxwell theory
(with constant $\phi$)
one must take the limit $z\rightarrow +\infty$ with $\theta$ finite, which makes a direct comparison to the
standard holographic superconductor (in which $AdS_2$ plays a crucial role) cumbersome.
To avoid some of these difficulties we will choose to work with a new radial coordinate $r$, in terms of which the IR in the zero temperature solution
is always located at $r=0$.
By performing the following transformation,
\begin{equation}
\begin{split}
\rho=r^{\frac{d}{2\theta-d z}}, \quad \rho_h=r_h^{\frac{d}{2\theta-d z}},\quad a_0=\tilde{a}_0,\quad \kappa=\frac{2\theta-d z}{d} \tilde{\kappa},\\
 z=\frac{2m-1}{m+n-1},\quad \theta=\frac{d(m-1)}{m+n-1},\quad  \tilde{L}^2=L^2 (m+n-1)^2 \, ,
\end{split}
\end{equation}
the finite temperature background solution takes the form
\begin{equation}\label{hyperscalmn}
\begin{split}
&ds^2_{\text{h.v.}}=-r^{2m} \tilde{f}(r) dt^2+\frac{\tilde{L}^2 dr^2}{r^{2m}\tilde{f}(r)}+r^{2n} d\vec{x}^{\,2},\\
& \tilde{f}(r)=1-\left(\frac{r_h}{r}\right)^{2m+dn-1},\quad A=\tilde{a}_0\, r^{2m+dn-1}\tilde{f}(r)\,dt,\quad \phi=\tilde{\kappa}\ln(r),\\
& \tilde{L}^2=\frac{(m+(d-1)n)(2m+dn-1)}{V_0},\quad \tilde{a}_0=\sqrt{\frac{2(m-n)}{Z_0(2m+dn-1)}},\quad \tilde{\kappa}^2=2 d n(1-n),\\
&\alpha=\frac{2(1-m-dn)}{\tilde{\kappa}},\quad \beta=\frac{2(1-m)}{\tilde{\kappa}} \, ,
\end{split}
\end{equation}
where $r_h$ denotes the location of the horizon. We have traded the scaling exponents $\{z, \theta\}$ for the two parameters\footnote{For
completeness we include the expression for $\{m,n\}$ in terms of the original exponents,
\beq
m = \frac{z d -\theta}{z d - 2\theta} \, , \qquad n = \frac{d -\theta}{z d - 2\theta} \, .
\eq}
$\{m,n\}$.
In terms of these, the
$AdS_2\times \mathbb{R}^d$ geometry\footnote{However, note that when $m=1$ and $n=0$ (i.e., the $AdS_2$ case) the above
transformation fails, because $(z, \theta, L)$ are not well defined.}
 is obtained by choosing $m=1,n=\alpha=\beta=0$, while
geometries that are conformal to $AdS_2\times \mathbb{R}^d$ correspond to $m+n=1$ with $m\neq 1, m\neq 1/2$.
Finally, in the extreme limit the metric and gauge field reduce to
\begin{equation}
\label{extreme}
ds^2_{\text{h.v.}}=-r^{2m} dt^2+\frac{\tilde{L}^2 dr^2}{r^{2m}}+r^{2n} d\vec{x}^{\,2},\quad A=\tilde{a}_0\, r^{2m+dn-1}\,dt \, ,
\end{equation}
with the scalar field maintaining its log form.
The temperature and entropy density associated with these black brane solutions are given by
\begin{eqnarray}\label{backT}
T=\frac{|2m+d n-1|}{4\pi \tilde{L}}\, r_h^{2m-1} \, ,\quad S=\frac{1}{4 G_N}r_h^{d n} \, ,
\end{eqnarray}
so that the thermal entropy can be seen to scale like\footnote{In the special case $m=1/2$, or equivalently $z=0$,
the temperature is independent of $r_h$.}
\begin{eqnarray}\label{app:entropy}
S \sim T^{\frac{dn}{2m-1}} \sim T^{\frac{d-\theta}{z}} \, ,
\end{eqnarray}
which can be interpreted as describing a system in which the degrees of freedom occupy an effective number of dimensions $\sim d_{eff} = d- \theta$.

In addition to the background geometry~\eqref{hyperscalmn}, in which the gauge field flux is non-trivial,
the theory we are considering admits another type of  hyperscaling violating solution for which $A_t$ vanishes identically,
given by
\begin{equation}
\label{hyperscalneu}
\begin{split}
& ds^2_{\text{h.v.}}=-r^{2n} \tilde{f}(r) dt^2+\frac{\tilde{L}^2 dr^2}{r^{2n}\tilde{f}(r)}+r^{2n} d\vec{x}^{\,2},\\
& \tilde{f}(r)=1-\left(\frac{r_h}{r}\right)^{(2+d)n-1},\quad A_t=0,\quad \phi=\tilde{\kappa}\ln(r),\\
& \tilde{L}^2=\frac{dn((2+d)n-1)}{V_0},\quad \tilde{\kappa}^2=2 d n(1-n),\quad \beta=\frac{2(1-n)}{\tilde{\kappa}},
\end{split}
\end{equation}
which can be considered as the special case of~\eqref{hyperscalmn} with $m=n$, or equivalently $z=1$.
These solutions are characterized entirely by the hyperscaling violating exponent $\theta$.
We will refer to them as IR neutral throughout the text.

\subsection{The asymptotic AdS solution}

To adopt the standard holographic dictionary and ensure a UV CFT, we would like to embed these solutions in AdS space.
This can be easily done by modifying the scalar potential $V(\phi)$ appropriately, so that the neutral field $\phi$
can settle to a constant value $\phi_{UV}$ at the boundary.
More specifically, the effective scalar potential $V_{eff} = V(\phi) + \frac{1}{4}Z(\phi)F^2$
will have to be chosen so that it admits an extremum at the UV fixed point,
$V^\prime_{eff}(\phi_{UV})=0$.
It will suffice to add a second exponential to (\ref{VandZ}), so that
 $$V(\phi) \rightarrow -V_0 e^{-\beta \phi} + V_1 e^{\gamma\phi} \, , $$ as done e.g. in~\cite{Iizuka:2011hg,Bhattacharya:2014dea}.
The transition scale $r_{tr}$ to AdS is then determined by the location at which the new term in the potential
begins to dominate over the original $V_0$ term.
The new exponential will then determine the properties of the AdS UV background solution.
In the numerical studies of Section \ref{NumericsSection} we will work for convenience with a scalar potential
\beq
V \sim \cosh\phi \, .\nn
\eq
However, more general choices can easily be implemented.
Furthemore, since we are interested in identifying the instabilities that arise solely from the hyperscaling violating
region of the geometry, any term in the potential which dominates only in the UV will not affect the main discussion of this paper.

\subsection{Constraints on the parameter space of the scaling exponents}
The allowed parameter space of the scaling exponents $\{m,n\}$ (or equivalently $\{z,\theta\}$),
can be restricted by imposing a number of physical constraints, which will ensure that the background can be taken to
describe a well-defined ground state.
Here we focus on the IR charged solution and exclude the $AdS_2$ geometry for the sake of greater clarity.
\begin{enumerate}[(a)]
\item
By inspecting the form of the metric note that in order for the solution~\eqref{hyperscalmn} to be real, we should demand
\begin{equation}\label{realsolu}
\tilde{L}^2>0,\quad \tilde{\kappa}^2>0,\quad \tilde{a}_0> 0\,,
\end{equation}
from which we obtain
\begin{equation}\label{irgood}
n(1-n)>0,\quad (m+(d-1)n)(2m+dn-1)>0, \quad \frac{m-n}{2m+dn-1}> 0 \, ,
\end{equation}
or alternatively in terms of $z$ and $\theta$,
\begin{equation}\label{realcondition}
(\theta-d)(\theta-d z+d)>0 \, , \quad (d-1+z-\theta)(d+z-\theta)>0 \, , \quad (z-1)(d+z-\theta)>0 \, .
\end{equation}
For the IR neutral case~\eqref{hyperscalneu}, the last relation must be set to zero, \emph{i.e.} $z=1$ or $m=n$.

\item
To have an unambiguous IR we should require the $(t, \vec{x})$
components of the metric scale in the same way with $r$ in~\eqref{extreme}, which means
\begin{equation}
\label{condb}
m\, n>0.
\end{equation}
The location of the IR depends on where the $(t, \vec{x})$ metric elements vanish.
Inspecting~\eqref{irgood} one finds that $m>0$ and $n>0$. Therefore the IR is located at $r=0$.

\item
To resolve the deep IR singularity of the geometry~\eqref{extreme}, we require the temperature deformation to be relevant,
following the discussion of~\cite{Gubser:2000nd,Charmousis:2010zz}.
This corresponds to the following constraint,
\begin{equation}\label{gubser}
2m+dn-1>0 \, , \quad \text{or equivalently} \quad \frac{d(z+d-\theta)}{zd-2\theta} > 0 \, ,
\end{equation}
which however is already imposed by (\ref{irgood}) and (\ref{condb}).

\item
We would like the geometry to have positive specific heat. From the scaling of the entropy with temperature, we should demand
\begin{equation}
\frac{dn}{2m-1}>0 \, , \quad \text{or equivalently} \quad \frac{d-\theta}{z} > 0 \, ,
\end{equation}
which implies that $m>\half$ since $n$ is positive.
\end{enumerate}

The allowed parameter range once we combine  all the conditions above is
given by\footnote{For the IR neutral background~\eqref{hyperscalneu}, the parameter range reads
\begin{equation}
\frac{1}{2}<(m=n)<1.
\end{equation}
}
\begin{equation}\label{rangemn}
\Bigl[ \, \frac{1}{2}<m\leqslant 1,\, 0<n<m\Bigr]\, ,\quad [m>1, \, 0<n<1]\, .
\end{equation}
For completeness we include the final $\{z,\theta\}$ parameter space in terms of the original $\rho$ coordinate used in (\ref{hyperscalz}),
\begin{equation}\label{rangeztheta}
\begin{split}
\text{IR located at}\quad \rho\rightarrow 0:\quad &[z<0, \theta>d]\,,\\
\text{IR located at}\quad \rho\rightarrow \infty:\quad&[1<z\leqslant 2, d+\theta<d z],\quad[z>2, \theta<d] \, .
\end{split}
\end{equation}
One can easily check that Null Energy Condition is automatically satisfied.

\section{Effective mass and superfluid instability windows}
\label{masssection}

Having introduced the properties of the background geometry we will be working with, we are now ready to examine
under what conditions the charged scalar field can condense.
For simplicity we will treat $\psi$ as a perturbation on top of the hyperscaling violating
solutions we have just discussed, and neglect the effects of backreaction.
Since we are zooming in on the transition point at which scalar hair begins to form -- the onset of the instability -- the $\psi$
scalar is going to be very small, and ignoring backreaction should be a good approximation.

In this section we are going to approach the question of instabilities by asking what we can learn from
the structure of the effective mass (\ref{meff}) of the charged scalar,
\begin{equation}
\label{meffSIV}
m^2_{eff} = - q^2 A_t^2 \, |g^{tt}| + B(\phi) \, ,
\end{equation}
and focus entirely on unstable modes which arise from the hyperscaling violating region of the geometry, $r_{IR}\leq r \leq r_{tr}$.
We will obtain simple analytical instability conditions which include, in the most tractable cases,
generalizations  of the well-known BF bound for AdS space.
Although we work for simplicity at zero temperature, we expect these conditions to capture all the essential features of
the finite temperature phase transition (as long as the temperature is not too large).
Indeed, this will be confirmed by the analysis of section \ref{NumericsSection}, where
we will revisit the intuition developed here by performing numerical studies in the background of finite temperature solutions.
Moreover, the same physics will be encoded in the
effective Schr\"{o}dinger potential analysis of the next section, where we will analyze these instability windows in greater detail.

As in the case of the standard holographic superconductor, a sufficiently large gauge field contribution will
drive (\ref{meffSIV}) negative, eventually causing the charged scalar field to condense.
The detailed properties of the condensate will be determined by the interplay between $B(\phi)$ and $A_\mu A^\mu$,
with the two contributions to the effective mass
competing against each other when $B$ is positive, and otherwise enhancing each other.
It will be the structure of the coupling $B(\phi)$ between the two scalars -- and in particular, how it scales compared to the $\{z,\theta\}$
background -- that will be at the root of the key differences with the standard AdS story.

Indeed, if we want to ensure that the mass term $B(\phi) \, \psi^2$ scales in the same way as the kinetic term
$(\p \psi)^2$, the coupling $B$ must be chosen appropriately\footnote{The additional coupling $C(\phi)$, which we set to unity,
would not affect the relative scaling between the kinetic and mass terms
but it would change the overall scaling of the term $ C(|D\Psi|^2 + B |\Psi|^2) $ in the action.}, as discussed \emph{e.g.} in~\cite{Lucas:2014zea}.
More precisely, in the hyperscaling violating portion of the geometry the kinetic term scales as
\begin{equation}
(\p \psi)^2 \sim   \tilde{f}(r) \, \frac{r^{2m}}{r^2} \, \psi^2 \sim r^{2(m-1)} \, \psi^2,
\end{equation}
where in the last expression we have switched off the temperature by taking $\tilde{f}(r)=1$.
Thus, in order for the $B \, \psi^2$ mass term to respect this scaling one needs
\begin{equation}
\label{Bscaling}
B(\phi) \sim r^{2(m-1)} \quad \quad \Rightarrow \quad \quad B (\phi)\sim e^{\frac{2(m-1)}{\tilde{\kappa}} \phi}.
\end{equation}
The gauge field contribution to the effective mass of the scalar, \emph{i.e.} $ q^2 A^2 \psi^2$, will generically scale
differently, in particular
\begin{equation}
q^2 \, A_\mu A^\mu \sim - q^2 \, r^{2(m+d n-1)} \, ,
\end{equation}
and will agree with (\ref{Bscaling}) only when $n=0$, or equivalently $\theta = d$,
which is outside the allowed parameter space ~\eqref{rangemn} and~\eqref{rangeztheta} of interest here\footnote{It may be worth examining this case
separately, as it could lead to a qualitatively different behavior.}.
In this paper we will take the coupling to be a generic power law (an exponential function of $\phi$),
\begin{equation}\label{scalB}
B(r) = B_0 \, r^\tau,
\end{equation}
with the case preserving the scaling of the kinetic term corresponding to
\begin{equation}\label{scaletau}
\tau=2 (m-1)\quad \Rightarrow\quad B=B_0 \, r^{2(m-1)} = B_0 \, r^{\frac{2\theta}{dz -2 \theta}} \, .
\end{equation}

It turns out to be convenient to let
$\psi(r)=r^{1-m- \half d n} g(r)$, so that the equation of motion for the charged scalar (\ref{chargedscalarEOM}) can be written
in the suggestive form
\begin{equation}\label{eomg}
\partial_r \left(r^2\partial_r g \right)=\left[\tilde{L}^2 \left(B_0 r^{\tau-2(m-1)}-Q^2 r^{2 dn}\right)+ \Bigl(m+\half d n\Bigr) \Bigl(m+\half d n-1\Bigr)
\right] g,
\end{equation}
where we have used~\eqref{scalB} and defined
\begin{equation}\label{effQ}
Q^2 \equiv \tilde{a}_0^2 q^2.
\end{equation}
Since the term on the left-hand side of  ~\eqref{eomg} is essentially the $AdS_2$ d'Alembertian (with the radius $L_{AdS_2}=1$),
we can interpret the right-hand side of the equation as defining the analog of an effective mass in $AdS_2$,  \emph{i.e.}
\begin{equation}\label{effM2}
M_{eff}^2(r) \equiv \tilde{L}^2[B_0 r^{\tau-2(m-1)}-Q^2 r^{2 dn}] + \Bigl(m+\half d n\Bigr) \Bigl(m+\half d n-1\Bigr) \, ,
\end{equation}
where the last constant term in terms of the original scaling exponents reads
\beq
\label{constant}
\Bigl(m+\half d n\Bigr) \Bigl(m+\half d n-1\Bigr) = \frac{(d^2-d\theta +2\theta)(d^2-d\theta +2zd -2d)}{4(zd-2\theta)^2} \, .
\eq
Indeed, if we set $Q=0$, $m=1$ and $n=\tau=0$, we recover the pure $AdS_2\times \mathbb{R}^d$ case,
for which $M_{eff}^2 = \tilde{L}^2 B_0$ and the solution to (\ref{eomg}) is
\beq
\label{ads2case}
g(r) \propto r^{-\half \pm \half \nu_0} \, , \qquad  \nu_0 = \sqrt{1+4\tilde{L}^2 B_0} \, ,
\eq
with the $AdS_2$ BF bound coming from requiring the index $\nu$ not to become imaginary,
\beq
B_0 \, \tilde{L}^2 \geq - \frac{1}{4} \, .
\eq

On the other hand, when $z\neq 1$, $\theta\neq 0$ and $Q \neq 0$ the effective mass (\ref{effM2})
depends generically on the radial coordinate,
and we lack a sharp local instability criterion, unlike in the simple $AdS$ case.
Still, instabilities can be expected to appear if $M_{eff}^2$ becomes negative enough.
Interestingly, even for generic values of the scaling exponents -- as long as they fall within the range~\eqref{rangemn} --
the constant term satisfies $\left(m+ \half d n \right) \left(m+ \half d n-1\right)>-\frac{1}{4}$ and remains above the $AdS_2$ BF bound.
Thus, it will be the combination of the charge term $\propto Q^2$ and the coupling $B$ which will typically
generate a sizable negative contribution to $M^2_{eff}$.
Indeed, it is apparent from (\ref{effM2})
that a scalar field condensate can form via two distinct mechanisms, as is well known.
First, a sufficiently large and negative $B(\phi)$ can trigger the transition, allowing even neutral scalars to condense (as already
known from AdS).
The second mechanism is the usual negative contribution to $M_{eff}^2$ coming from the gauge field term,
which can make it energetically favorable for the charged scalar to condense.

However, there are some key differences with the usual holographic superconductor setup.
First, depending on the scaling behavior of $B(\phi)$ interesting competitions between the two mechanisms can be generated.
More importantly, note that the contribution to (\ref{effM2}) from the $U(1)$ gauge field becomes less and less important
as the IR is approached\footnote{In the $AdS_2$ case $n=0$, therefore the $U(1)$ gauge field has a finite contribution in the IR.}, and
vanishes at $r = 0$.
Thus, here we expect instabilities associated with the charge term to be generically localized close to $r_{tr}$
(or possibly at some intermediate radial distance $r$ at which the effective mass $M_{eff}^2(r)$ has a deep negative minimum) and not in the IR.
As we will see shortly, this behavior can be modified for certain choices of $B$, but it is otherwise robust.
Below we are going to make the discussion more quantitative by highlighting a few cases, and leave a more detailed analysis to
Section \ref{SchrodingerSection}.

\subsection{Scaling case $\tau=2(m-1)$}

\noindent
(i) {\bf \underline{Neutral scalar}:} \\
\noindent
We consider first the case of a neutral scalar. When $Q=0$ the effective mass is just a constant,
\begin{equation}\label{llneutralmass}
M_{eff}^2=\tilde{L}^2B_0+\Bigl(m+\half d n\Bigr) \Bigl(m+\half d n-1\Bigr)\, ,
\end{equation}
and the $\psi$ perturbation has the power law form
\beq\label{exp1}
\psi(r) = r^{1-m-\half d n} \, g(r) = r^{\half-m-\half \, d \, n \pm \half \, \nu} = r^{-\half \frac{d(z+d-\theta)}{zd-2\theta} \pm \half \, \nu} \, ,
\eq
with the exponent $\nu$ given by
\bea
\label{nuexp}
\nu &=& \sqrt{1+4M_{eff}^2} = \sqrt{4\tilde{L}^2B_0 + \frac{d^2 (z+d-\theta)^2}{(dz-2\theta)^2}} \nn\\
&=& \frac{\tilde{L}}{L}\sqrt{4 B_0 L^2 + (z+d-\theta)^2} \, .
\ea
Requiring the index $\nu$ not to become imaginary
immediately leads to the non-relativistic, hyperscaling violating analog of the standard AdS BF bound,
\beq
\label{generBF1}
4 B_0 L^2 \geq - (z+d-\theta)^2  \, .
\eq
It can be equivalently expressed in terms of the effective mass,
\beq
\label{generBF2}
M_{eff}^2 \geq - \frac{1}{4} \,,
\eq
which interestingly tells us that the onset of the instability is controlled by $M_{eff}^2$ dipping below
the critical mass saturating the $AdS_2$ BF bound, $M^2_{AdS_2}= -1/4$, even with generic scaling exponents
$z\neq 1$, $\theta\neq0$.
Thus, in these scaling backgrounds we expect a \emph{neutral} scalar to be able to condense provided the value of $B_0$
is negative enough to violate (\ref{generBF2}), as in the simpler AdS case.
Finally, we note that the generalized BF bound (\ref{generBF1}) was already obtained in \cite{Gath:2012pg}.\\

\noindent
(ii) {\bf \underline{Charged scalar}:}\\
\noindent
When we restore the charge, the effective mass becomes radially dependent,
\begin{equation}
M_{eff}^2=\tilde{L}^2 B_0-\tilde{L}^2 Q^2 r^{2 dn}+\Bigl(m+\half d n\Bigr) \Bigl(m+\half d n-1\Bigr) \, ,
\end{equation}
and the perturbation $g(r)$ is a combination of Bessel functions,
\beq\label{scalcharge}
g(r) = c_1 J_{\tilde{\nu}} \left(\frac{Q\tilde{L}}{dn} r^{dn}\right) + c_2 Y_{\tilde{\nu}} \left(\frac{Q\tilde{L}}{dn} r^{dn}\right) \, ,
\eq
with the index $\tilde{\nu}$ of the Bessel functions related to that appearing in (\ref{nuexp}) through
\beq
\tilde{\nu} = \frac{1}{2dn} \, \nu     \, .
\eq
Thus, as in the case of vanishing charge, there will be an instability when the mass term $\sim B_0$ is so negative that it violates
the generalized BF bound (\ref{generBF2}), corresponding to the index of the Bessel functions becoming imaginary.

Of course, there is an additional source of instability which is driven by the charge term becoming sufficiently negative.
Unlike in the case of the standard holographic superconductor~\cite{Hartnoll:2008kx}, however, here the gauge field term (which approaches zero
as $r \rightarrow 0$) dominates not in the deep IR, but rather near the $r \sim r_{tr}$ transition region to AdS.
Indeed, within the hyperscaling violating portion of the geometry,
$Q^2 r^{2dn}$ attains its largest value at $r=r_{tr}$, and that is where
we expect the superfluid instability to be localized.
As a result, a \emph{necessary} condition for the formation of instabilities is
\beq
\tilde{L}^2 Q^2 r_{tr}^{2dn} > \tilde{L}^2 B_0 +  \Bigl(m+\half d n\Bigr) \Bigl(m+\half d n-1\Bigr) \, ,
\eq
which can be satisfied by increasing the charge or alternatively pushing the transition region $r_{tr}$ closer and closer to the UV.
Note that in these constructions $r_{tr}$ plays a crucial role in controlling the onset of the phase transition. 
The discussion above breaks down in the IR neutral background~\eqref{hyperscalneu}, 
for which $Q=0$ while $q\neq 0$. 
The effective mass $M_{eff}^2$ is the same as that of the neutral case~\eqref{llneutralmass} but with $z=1$, 
and therefore~\eqref{generBF2} is the appropriate criterion for the superfluid instability triggered in the IR. 
We will not stress this special case in what follows.

Finally, to describe $AdS_2\times \mathbb{R}^d$ with $Q \neq 0$, we set $m=1, n=\tau=0$ to find
\begin{equation}
M_{eff}^2 =\tilde{L}^2B_0-\tilde{L}^2 Q^2 \, ,
\end{equation}
leading to the well-known $AdS_2$ instability window
\begin{equation}
\label{ads2BF}
M_{eff}^2=\tilde{L}^2B_0-\tilde{L}^2 Q^2 <- \frac{1}{4} \, .
\end{equation}

\subsection{Non-scaling case, $\tau\neq 2(m-1)$}
\label{nonscalneg}

\noindent
(i) {\bf Parameter choices $\tau<2(m-1)$ and $B_0<0$:}

\noindent
The coupling $B(\phi)$ contribution to~\eqref{effM2} approaches negative infinity as $r\rightarrow 0$,
while the remaining terms in~\eqref{effM2} stay finite.
Compared to the scaling case, the effective mass here is much more negative along radial flow towards the IR, and thus
instabilities are expected to be generic and form much more easily.
Moreover, there should be unstable modes at arbitrarily small values of the charge $Q$,
associated with the deep IR portion of the geometry.
As a consequence, we expect neutral scalars to condense generically, independently of how small or large $B_0$ is (in contrast to the standard
AdS case).
We will return to this point in the next section, but anticipate to be able to find a superfluid
phase transition at arbitrarily low temperature and charge. \\

\noindent(ii) {\bf Parameter choices $\tau<2(m-1)$ and $B_0>0$:}\\
\noindent
On the other hand, in this case the contribution to~\eqref{effM2} coming from the coupling $B$ will approach
positive infinity as $r\rightarrow 0$, preventing the formation of an unstable mode in the deep IR.
Nevertheless, a sufficiently large value of the charge $Q$ may trigger a superfluid instability near the scale $r\sim r_{tr}$,
where the gauge field term $\sim Q^2 r^{2dn}$ is largest.
For this parameter range we expect that a minimal charge will be needed in order for the charged condensate to form.
We will examine this point in detail in Section \ref{SchrodingerSection}.\\

\noindent(iii) {\bf Parameter choices $\tau>2(m-1)$:}\\
\noindent
When $\tau>2(m-1)$, the two terms $B_0 r^{\tau-2(m-1)}$ and $Q^2 r^{2 dn}$ in~\eqref{effM2} both vanish at $r\rightarrow 0$, and it is
challenging to obtain a clean instability criterion. Whether an unstable mode will be present depends
on whether the $Q$ and $B_0$ terms will compete against each other (when $B_0>0$) or enhance each other (when $B_0<0$).
Generically we expect to find a minimal charge $Q_{min}$ below which no instabilities will form.
It is difficult to be more quantitative at this stage, but we will return to these two cases in more detail in \ref{SchrodingerSection}.\\

\noindent(iv) {\bf Parameter choices $\tau-2(m-1)=2dn$:}\\
\noindent
When $\tau = 2(m-1) + 2dn$ we see that the coupling and gauge field terms in~\eqref{effM2} scale in the same way,
$\propto r^{2dn} \left[ B_0 - Q^2 \right]$. Thus, when $B_0 > Q^2$ the effective mass will never be negative 
in the hyperscaling violating portion of the geometry, ensuring the absence of
instabilities in that regime. Interestingly, this is true even for very large charge.
On the other hand, when $B_0 <Q^2$ the radially dependent part of $M_{eff}^2$ will be negative, but will approach
zero towards the IR. Thus, we expect to have a condensate as long as the effective mass can become sufficiently negative near $r_{tr}$.
However, even in this case we will always have a minimal charge, since $r^{2dn} \rightarrow 0$ towards the deep IR and the constant term in the
effective mass is positive. Note that the superfluid instability can seemingly be triggered by varying the
coupling across the critical value $B_0 = Q^2$.
We leave a more detailed treatment of this case to future work.

We are now ready to compare the intuition developed here with what one can learn by recasting the scalar equation
in the form of a Schr\"{o}dinger equation.
Indeed, the presence of bound states in the Schr\"{o}dinger potential can also be taken as an indicator of instabilities, as we discuss next.

\section{Effective Schr\"odinger Potential and Instabilities}
\label{SchrodingerSection}

By an appropriate combination of a change of coordinates and a field redefinition,
the charged scalar field equation of motion (\ref{chargedscalarEOM}) can be rewritten in Schr\"odinger form
-- as done, for example, in holographic studies of the conductivity~\cite{Horowitz:2009ij}.
Inspecting the sign of the resulting Schr\"odinger potential can then offer a window into the
presence of instabilities in the system \cite{Hartnoll:2008kx}.
In particular, if the Schr\"odinger equation has a negative energy bound state,
there will be unstable modes.
Negative energy in this context corresponds to $\omega^2<0$, \emph{i.e.} imaginary frequencies and
therefore solutions which grow exponentially in time.
Also, if for a certain range of parameters the effective potential remains positive \emph{everywhere} in the hyperscaling
violating portion of the geometry, we are guaranteed the absence of superfluid instabilities there.

We turn on the charge of $\psi$ and work with the parametrisation given by~\eqref{extreme},
taking the coupling to the neutral scalar to be  $B=B_0 \, r^{\tau}$.
Recall that $\tau=2(m-1)$ is the case that preserves some of the scaling symmetry.
We work at zero momentum and take $\psi = e^{- i\omega t}\, \psi(r)$.
Introducing a new radial variable $\xi$ and rescaling the charged scalar field,
\begin{equation}
\frac{d\xi}{d r}=\tilde{L} r^{-2m},\quad \tilde{\psi}(\xi)=r^{dn/2}\, \psi(r),
\end{equation}
the equation for the perturbation takes the form of a Schr\"odinger equation
\begin{equation}
-\frac{d^2}{d\xi^2}\tilde{\psi}+ V_{Schr}\,\tilde{\psi}=\omega^2\, \tilde{\psi} \, ,
\end{equation}
with the effective Schr\"odinger potential given by
\begin{equation}\label{schrpoten}
\begin{split}
\tilde{L}^2 V_{Schr}(r)&=r^{2(2m-1)}\left[\tilde{L}^2 \left(B_0 r^{\tau-2(m-1)}-Q^2 r^{2 d n}\right)+\frac{dn(dn+4m-2)}{4}\right],\\
\end{split}
\end{equation}
where we used the original radial coordinate for simplicity and we recall that $Q$ was defined in~\eqref{effQ}.
Notice that the overall factor $r^{2(2m-1)}\rightarrow 0$ in the far IR, as $r\rightarrow 0$.

The last, constant term in the potential happens to be positive definite, as
one can see from~\eqref{rangemn}, and can be repackaged in the following form,
\beq
dn(dn+4m-2) = \frac{\tilde{L}^2}{L^2} \Bigl[(z+d-\theta)^2 - z^2 \Bigr] =
\nu^2 - 4 B_0 \tilde{L}^2 - z^2  \frac{\tilde{L}^2}{L^2} >0 \, ,
\eq
where $\nu$ was introduced in (\ref{nuexp}). This expression can be used to rewrite
 the potential in the following suggestive form,
\beq
\label{Vschr2}
V_{Schr}(r) =r^{2(2m-1)}\left[ B_0 r^{\tau-2(m-1)}-Q^2 r^{2 d n} +
\frac{1}{4} \left( \frac{\nu^2}{\tilde{L}^2} - 4 B_0 - \frac{z^2}{L^2} \right) \right] \, .
\eq

To recap, unstable modes will correspond to negative energy bound states for which $\omega^2<0$,
with the critical case describing zero modes associated with $\omega^2=0$.
Thus, a \emph{necessary} condition for the existence of an instability is that the potential $V_{Schr}$ develops
at least one negative region in the bulk.\footnote{The statement can be made more precise if we know the profile of the
potential~\eqref{schrpoten} in the entire bulk region, from the IR to the UV. By using the WKB approximation, one can obtain a
bound state at zero energy in a potential well for each integer $k\geqslant 1$ via the formula
\begin{equation}\label{wkb}
(2k-1)\pi=2\int d\xi \sqrt{-V_{Schr}(\xi)},
\end{equation}
where the integral is carried out in the region of negative Schr\"{o}dinger potential.
We leave the study of this interesting feature to future work.}
Inspecting (\ref{Vschr2}) we see that the possible sources of instabilities are again
transparent: the relative interplay between the charge, the coupling $B$ and the value of the index $\nu$.
Recall that in this paper we are only after unstable modes associated with the
hyperscaling violating region itself~\footnote{The AdS UV geometry may have additional instabilities
which are not captured by the behaviour of (\ref{schrpoten}). However, those are already well understood
via standard BF bound arguments, and will be ignored here.},
for which $0\leq r \leq r_{tr}$ when the IR is at $r =0$. As a consequence,
we are only looking for negative regions of \eqref{Vschr2}, and not of the potential which determines the
UV behavior of the theory and the asymptotic AdS geometry.

\subsection{Neutral Scalar}

Let's focus on the neutral scalar case $Q=0$ first, and consider different choices for the coupling:
\begin{enumerate}
\item
{\bf Scaling choice $\tau=2(m-1)$:}\\
When the effective mass term $\sim B(\phi) \psi^2$ respects the scaling of the kinetic term,
the potential reduces to the simple expression
\beq
\label{Vneutralscaling}
V_{Schr}(r) = r^{2(2m-1)}\left[B_0 + \frac{1}{4 \, L^2} \left[ (z+d-\theta)^2-z^2 \right] \right]
= r^{2(2m-1)} \frac{1}{4} \left[ \frac{\nu^2}{\tilde{L}^2}  - \frac{z^2}{L^2} \right] \, ,
\eq
and is always positive everywhere in the hyperscaling violating part of the geometry if  $B_0>0$.
Thus, to trigger any instabilities one necessarily needs to have $B_0<0$.

In terms of the index $\nu$, the condition for $V_{Schr}<0$ is $\nu^2 < z^2 \tilde{L}^2/L^2$.
Notice however that the violation of the generalized BF bound (\ref{generBF1}) corresponds to a smaller window,
\beq
\nu^2 <0 \, ,
\eq
associated with the index becoming imaginary.
Thus, we see an offset\footnote{An explanation for the origin of the shift $\propto z^2/4$
was provided by \cite{Keeler:2013msa}.
We thank Jim Liu for bringing this to our attention.}
(by an amount $\propto \, z^2$) between the violation of the generalized BF bound and the condition $V_{Schr}\leq0$.
However, one should keep in mind that $V_{Schr}\leq0$ is \emph{not}
a sufficient condition for instabilities, but only a necessary one.
In other words, the potential should be ``negative enough'' in order for bound states to form,
and one should quantify how deep the potential well needs to be.

One way to test whether in the additional window
\beq
\label{window1}
0 < \frac{\nu^2}{\tilde{L}^2} < \frac{z^2}{L^2} \, ,
\eq
the $\psi$ scalar may condense (without a violation of the BF bound) is to examine the behavior of its IR perturbations.
In particular, in order for a condensate to form we must have at least one irrelevant perturbation mode in the IR,
without which a non-trivial scalar profile would not be supported\footnote{If the perturbations of $\psi$ were relevant, we would expect backreaction 
of the charged scalar on the background
to become important, and to lead to a new geometry which would not be that of our simple $\{z,\theta\}$ solutions. While this situation
is clearly interesting, it is beyond the scope of our paper, and we will not consider it here.}.
Indeed, recall that in Section \ref{masssection} we found that in the IR the scalar
had the form (\ref{exp1}), with modes
\beq
\psi \sim r^{-\half \frac{d(z+d-\theta)}{zd-2\theta} \pm \half \, \nu} \, ,
\qquad \nu^2 = 4\tilde{L}^2B_0 + \frac{d^2 (z+d-\theta)^2}{(dz-2\theta)^2} \, .
\eq
Since $\frac{d(z+d-\theta)}{zd-2\theta}>0$,  as seen from~\eqref{gubser}, and the IR corresponds to $r=0$,
one can easily check that the range (\ref{window1}) does \emph{not} allow for irrelevant perturbations\footnote{There are no IR
irrelevant modes in the larger window $0<\nu^2< \frac{\tilde{L}^2}{L^2}(z+d-\theta)^2$.
Notice that  $z^2 < (z+d-\theta)^2$ in our parameter space~\eqref{rangeztheta}.} and hence is ruled out as a possible ``condensation window".

The precise windows of instability in a given model can of course be tested numerically, using these analytical arguments as guidance.
We will return to this issue in Section \ref{NumericsSection}, but
for now let's summarize by pointing out that we have identified two mechanisms that will indicate the presence of a condensate.
First, the violation of the analog of the AdS BF bound. Second, the presence of IR irrelevant modes, without which
the boundary conditions which would allow for a condensate would not be satisfied.
Thus, the absence of irrelevant modes for the IR expansion of the neutral or charged scalar can be used as a criterion against
condensation in certain regions of parameter space, especially in cases for which the Schr\"{o}dinger potential
analysis is not necessarily conclusive.

\item {\bf Arbitrary scaling $B = B_0 \, r^\tau$:}\\
The Schr\"{o}dinger potential is now given by
\beq
V_{Schr}(r)=r^{2(2m-1)}\left[B_0 \, r^{\tau-2(m-1)} + \frac{1}{4 \, L^2} \left[ (z+d-\theta)^2-z^2 \right] \right] \, .
\eq
Again, to trigger any instabilities one needs $B_0 <0$ and negative enough to overcome the positive contribution of the constant term.
Thus, take $B_0<0$ and consider the two cases:
\begin{enumerate}[(i)]
\item
Let's assume first that $\tau - 2(m-1) >0$, so that the coupling $B(\phi)$ approaches zero towards the IR.
Then, in the hyperscaling violating portion of the geometry the term $|B_0| \, r^{\tau - 2(m-1)}$ is largest when $r =r_{tr}$.
This implies that we are guaranteed no instabilities when
\beq
|B_0| \, r_{tr}^{\tau - 2(m-1)} < \frac{1}{4 \, L^2} \left[ (z+d-\theta)^2-z^2 \right] \, ,
\eq
since the potential is, again, everywhere positive in that case.

\item
On the other hand, when $\tau - 2(m-1) <0$ so that the coupling is becoming increasingly negative towards the IR,
instabilities are expected to arise quite generically, and to be associated with the IR of the geometry.
\end{enumerate}

\end{enumerate}

\subsection{Charged Scalar}

As can be easily seen from (\ref{schrpoten}), since $n>0$ the charge contribution to the potential $V_{Schr}$ always decreases
towards $r=0$, and is therefore largest precisely near the transition region $r\sim r_{tr}$ to AdS.
As a result, we expect the bound states to be generically\footnote{This will be the case when the coupling $B(\phi)$ is of the scaling form.
On the other hand, when $B$ is chosen to diverge towards the IR, and $B_0<0$, this story will change, as we will see.}
localized there and not in the deep IR.
This is in sharp contrast with the standard holographic superconductor setup
with an $AdS_2$ IR region, for which the charge contribution $\sim Q^2 r^{2dn}$ is constant, as $n=0$.
Once again, we are going to examine the structure of the effective Schr\"{o}dinger potential at zero temperature
for different choices of coupling $B(\phi)$, but this time with $Q \neq 0$:
\begin{enumerate}
\item
{\bf Scaling choice $\tau=2(m-1)$:}\\
\noindent
In the presence of charge we have
\beq\label{potentialQ}
V_{Schr}(r) =r^{2(2m-1)}\left[- Q^2 r^{2 d n} +
\frac{1}{4} \left( \frac{\nu^2}{\tilde{L}^2} - \frac{z^2}{L^2} \right) \right]\, .
\eq
We consider the following cases:

(i) When $\frac{\nu^2}{\tilde{L}^2} <   \frac{z^2}{L^2}$
the effective Schr\"{o}dinger potential
is negative everywhere independently of how large the charge is. Thus, we expect the scalar to be
able to condense even when $Q$ is very small\footnote{This was already anticipated by the neutral scalar analysis discussed above.}.
In particular, when the stricter condition
\beq
\nu^2 <0\,,
\eq
is satisfied,
the condensation is triggered at zero charge, as anticipated by the neutral scalar
field analysis above.  Note that this particular neutral scalar field instability -- which is nothing but the violation of the generalized BF bound --
originates from the far IR of the geometry. It is visible both from the behavior of the effective mass as well as
from the Schr\"{o}dinger potential~\eqref{potentialQ}.

On the other hand, when $0<\nu^2<z^2  \frac{\tilde{L}^2}{L^2} \,$,
even though the Schr\"{o}dinger potential~\eqref{potentialQ} develops a negative region as $r\rightarrow 0$,
we are not guaranteed the onset of a superfluid phase transition in the far IR.
Indeed, recall that in this range the IR perturbations of a neutral field are inconsistent with the formation of a condensate -- there are no
IR irrelevant perturbations.
A similar perturbation analysis needs to be done for the charged scalar, to ensure that the IR mode expansion is compatible
with the presence of a condensate.
Indeed, from~\eqref{scalcharge}, we can obtain the asymptotic behavior in the far IR,
\begin{equation}
\psi= r^{-\half \frac{d(z+d-\theta)}{zd-2\theta}- \half  \nu}(c_1+Q^2 r^{2dn}+\mathcal{O}(Q^4 r^{4dn}))+
r^{-\half \frac{d(z+d-\theta)}{zd-2\theta}+\half  \nu}(c_2+Q^2 r^{2dn}+\mathcal{O}(Q^4 r^{4dn})),
\end{equation}
with $\nu^2 = 4\tilde{L}^2B_0 + \frac{d^2 (z+d-\theta)^2}{(dz-2\theta)^2}$.
Notice that the contribution from $U(1)$ gauge field only appears as subleading corrections.
Once again, one can easily see that the range $0<\nu^2<z^2  \frac{\tilde{L}^2}{L^2}$ does not allow for any irrelevant mode and
is therefore ruled out as a viable condensation window.
Thus, we have seen explicitly that having a negative
region in the effective potential is not enough to trigger an instability -- it is only a necessary condition, as we have stressed at the beginning.

Finally, since the charge term in the brackets of~\eqref{potentialQ} contributes more and more as we move away from the IR
while the coupling $B$ doesn't scale, bound states of the potential will typically be supported near
$r_{tr}$, for a large enough value of $Q$. Thus, superfluid instabilities of the hyperscaling violating regime
will be associated with the ``effective UV'' of the $\{z,\theta\}$ geometry itself, and not with its IR region.
It is the dependence of the gauge field term on the hyperscaling violating exponent which is responsible for this
behavior, as visible from the structure of the Schr\"{o}dinger potential.

(ii) When $\frac{\nu^2}{\tilde{L}^2} >   \frac{z^2}{L^2}$, the potential will be positive at least in the far IR,
where the gauge field term becomes negligible independently of how large the charge is.
Whether $V_{Schr}$ can become negative in a different portion of the geometry depends on the interplay between $B_0, Q$, the scaling exponents and
the location of the transition region to AdS.

In particular, if the charge and transition region obey
\beq
Q^2 r_{tr}^{2dn} < \frac{1}{4}\left[\nu^2 - z^2  \frac{\tilde{L}^2}{L^2}  \right] \, ,
\eq
we are guaranteed the absence of unstable modes in the hyperscaling violating regime, since the Schr\"{o}dinger potential in this case is positive
in the entire bulk region (the charge term attains its largest value at $r_{tr}$).
While this condition can be easily evaded by increasing $Q$, it does translate into the existence of a \emph{minimal charge} $Q_{min}$
below which the superfluid phase transition can not be triggered.
In particular,
\beq
\label{Qminimal1}
Q > Q_{min} \, , \qquad  \text{with} \qquad
Q_{min}^2 \equiv \frac{1}{r_{tr}^{2dn}}\frac{1}{4}\left[\nu^2 - z^2  \frac{\tilde{L}^2}{L^2}  \right]  \, ,
 \eq
is \emph{a necessary condition} for the existence of instabilities.
Note that the the minimal charge can be made smaller by increasing $r_{tr}$, \emph{i.e.} the range in which the
hyperscaling violating solution dominates the geometry, or alternatively by increasing $n=\frac{d-\theta}{zd - 2\theta}$.

The presence of a minimal charge when the mass $\sim B_0$ of a charged scalar is either positive or ``not negative enough'' is
by no means new, and is well known to occur in AdS. In fact, it is already encoded in the physics of the generalized BF bounds we described in
Section~\ref{masssection}. In this respect this case is analogous to what happens in the standard holographic superconductor setup.
We will see shortly that this story is modified when we allow for more general scalings $B(\phi)$.

It is hard to make more definite statements about the precise onset of instabilities from the Schr\"{o}dinger potential alone.
One robust feature we already emphasized is that the condensate should be triggered close to the transition scale to AdS,
and not in the deep IR. As long as the charge is above $Q_{min}$, some portion of the hyperscaling violating geometry will correspond
to a negative potential, and we expect the charged scalar to condense.
However, we can't predict how negative the potential must be to support an unstable mode.

\item
{\bf Arbitrary scaling $B = B_0 \, r^\tau$:}\\
\noindent
The potential has the form
\beq
V_{Schr}(r) =r^{2(2m-1)} \left[ B_0 r^{\tau-2(m-1)}-Q^2 r^{2 d n}+
\frac{1}{4\,L^2} \left[(z+d-\theta)^2 - z^2 \right] \right] \, .
\eq
We distinguish between two different cases, depending on the sign of $B_0$: 
\begin{enumerate}[(i)]
\item
When $B_0>0$ the only negative contribution to the potential is from the charge term, which always decreases in magnitude towards the IR.
While this implies generically the existence of a minimal charge, what sets
its value depends on whether $\tau$ is larger or smaller than the scaling choice $2(m-1)$:
\begin{enumerate}[(a)]
\item
Let's consider first $\tau<2(m-1)$.
If a given charge is not large enough to make the potential negative at the transition scale,
it certainly has no chance of achieving it closer to the IR, because the coupling term $\propto B_0$ will only increase as $r\rightarrow 0$,
while the charge contribution will become weaker.
Thus, the potential is guaranteed to be positive along the entire region $0<r<r_{tr}$.
This tells us that there will be a minimal charge below which the condensate will not form, set by $r=r_{tr}$,
\beq
\label{Qmincond}
Q_{min}^2 = r_{tr}^{2dn} \left[ B_0 \, r_{tr}^{\tau-2(m-1)} + \frac{1}{4\,L^2} \left[(z+d-\theta)^2 - z^2 \right] \right] \, .
\eq
We can adjust $Q_{min}$ by varying the size $r_{tr}$ of the hyperscaling violating regime (the larger the region, the smaller the minimal charge),
as well as by increasing $n=\frac{d-\theta}{zd - 2\theta}$.

\item
The situation for $\tau>2(m-1)$ is more complicated, and one has to take into account the relative scaling between the charge and the coupling terms
to identify what sets the value of $Q_{min}$. The existence of a minimal charge is still generic because, as the IR is approached, at some point
the positive constant term will dominate the potential, unless the charge is increased above some critical value.

The special value $\tau = 2(m-1) + 2dn$ discussed in case (iv) of Section \ref{nonscalneg} naively falls within this category, but needs to be treated
separately. Indeed, notice that when $B_0>Q^2$ the potential is always positive, no matter how large the charge is.
Thus, a condensate will not form.
\end{enumerate}

\item
When $B_0<0$, we can rewrite the potential suggestively as
\beq
V_{Schr}(r) = r^{2(2m-1)} \left(\frac{1}{4\,L^2} \left[(z+d-\theta)^2 - z^2 \right] - \left[ |B_0| r^{\tau-2(m-1)}+Q^2 r^{2 d n}\right]  \right) \, .
\eq
Again the behavior depends on the range of $\tau$:
\begin{enumerate}[(a)]
\item
When $\tau>2(m-1)$, the potential in the deep IR is also positive, since the last two terms are approaching zero while the first constant term is positive.
Notice that this is true independently of how big $Q$ and $B_0$ are taken to be, which is different from the scaling choice,
in which one can simply tune $B_0$ to be large enough to trigger the instability in the deep IR.
Thus, to find $V_{Schr}<0$ one must approach the effective UV of the hyperscaling violating regime.
Once again, the largest the term $|B_0| r^{\tau-2(m-1)}+Q^2 r^{2 d n}$ can be is
set by the transition scale to AdS, $r= r_{tr}$. Thus, superfluid instabilities will not develop as long as
\beq
|B_0| r_{tr}^{\tau-2(m-1)}+Q^2 r_{tr}^{2 d n} < (z+d-\theta)^2 - z^2 \, .
\eq
This again sets a minimal charge, as in the previous cases. The new feature compared to the standard
holographic superconductor~\cite{Hartnoll:2008kx} is that the minimal charge is present independently of
how negative $B_0$ is tuned to be.
The special choice $\tau = 2(m-1) + 2dn$ discussed in case (iv) of Section \ref{nonscalneg} falls within this category.

\item
On the other hand, when $\tau<2(m-1)$ the contribution from the coupling $B$ becomes infinitely negative in the deep IR.
As a result, we expect to have a charged scalar condensate (this time localized in the far IR) for any value of the charge,
no matter how small it is, without having to tune $B_0$ to be large.
This is in contrast to the AdS case, for which the instability is associated with the mass term $m^2$ being very large and negative.
This is a new feature, due entirely to having allowed for an arbitrary scaling for $B$.
\end{enumerate}

\end{enumerate}

\end{enumerate}

To summarize, the simple analytical arguments we have formulated can be used to highlight the competition between
different sources of instabilities -- in particular, the interplay between the coupling $B(\phi)$ and the charge term --
and the criteria under which they are triggered or suppressed.
Although the analysis was performed using extremal solutions, it provides guidance to detailed
numerical studies of instability windows, and analytical intuition for when a minimal charge should exist.
Next, we will examine our estimates numerically.

\section{Numerics}
\label{NumericsSection}

So far our discussion has been restricted to zero temperature solutions,
but in what follows we will switch on a finite temperature, and examine these instability windows in the background of
hyperscaling violating black branes that are asymptotic to AdS.
For simplicity, we are going to work with an analytical solution which arises from the supergravity setup of~\cite{Gubser:2009qt}, and
is characterized by $z,\theta \rightarrow \infty$ with the ratio $\theta/z$ held fixed.
We will examine the condensation of the charged scalar on top of this analytical background numerically, in a number of examples
which will lend evidence to the simple estimates of the last two sections.
Although the latter apply only to extremal solutions, they provide a guide towards a classification of superfluid transitions at finite (but low)
temperature.
Finally, even though the $z,\theta \rightarrow \infty$ limit is rather special, we believe that it captures all the essential
features of our analysis, and postpone a more thorough look at black brane solutions with finite $z$ and $\theta$ to future work.

Working with $d=2$ and choosing the scalar potential and gauge kinetic function in~\eqref{Lag} to be
\begin{equation}
Z(\phi)=e^{\phi/\sqrt{3}} \, ,\quad V(\phi)=-6\cosh(\phi/\sqrt{3})\, ,
\end{equation}
we obtain the three-equal-charge black brane solution of~\cite{Gubser:2009qt},
\begin{equation}\label{3charge}
\begin{split}
&ds^2  = - f(r) dt^2+\frac{1}{f(r)}dr^2+h(r)(dx^2+dy^2),\\
&f= r^{1/2} (r+Q)^{3/2}\left(1-\frac{(r_h+Q)^3}{(r+Q)^3}\right),\quad h=r^{1/2} (r+Q)^{3/2},\\
&A_t=\sqrt{3Q(r_h+Q)}\left(1-\frac{r_h+Q}{r+Q}\right),\quad \phi=\frac{\sqrt{3}}{2}\ln(1+Q/r) \, ,
\end{split}
\end{equation}
where $r_h$ denotes the horizon.
The corresponding temperature and chemical potential are given by
\begin{equation}
T=\frac{3\sqrt{r_h (r_h+Q)}}{4\pi}\, ,\quad \mu=\sqrt{3 Q (r_h+Q)}\, .
\end{equation}
The extreme limit $T/\mu\rightarrow 0$ is obtained by taking $r_h/Q=0$, and the corresponding IR geometry has the
hyperscaling violating form~\eqref{extreme} with exponents $m=3/4$ and $n=1/4$.
Note\footnote{This case is known as semi-local criticality
and can give rise to interesting behavior~\cite{Hartnoll:2012wm,Donos:2012ra,Anantua:2012nj,Donos:2012yi}.}
that this solution is conformal to $AdS_2\times \mathbb{R}^2$.
More precisely, if we  introduce $\rho=\sqrt{\frac{Q}{3r}}$, then the extreme IR limit of~\eqref{3charge} can be written as
\begin{equation}\label{ir3charge}
\begin{split}
&ds^2  = \frac{Q^2}{\sqrt{3}}\frac{1}{\rho}\left[- \frac{dt^2}{\rho^2}+\frac{4}{3 Q^2}\frac{d\rho^2}{\rho^2}+dx^2+dy^2\right],\\
&A_t=\frac{Q}{\sqrt{3}}\frac{1}{\rho^2},\quad\quad \phi=\sqrt{3}\ln(\sqrt{3}\rho) \, ,
\end{split}
\end{equation}
with the IR now located at $\rho\rightarrow\infty$.
This kind of geometry can be obtained from~\eqref{hsvzeroz} by considering the limit $z, \theta \rightarrow \infty$ with $\theta/z=-1$.

To study the onset of superfluid instabilities in the background~\eqref{3charge},
we turn on a fluctuation $\psi=e^{-i\omega t}\psi(r)$ of the charged scalar, whose linearized
equation of motion reads
\begin{equation}\label{mode}
\psi''(r)+\left(\frac{f'}{f}+\frac{h'}{h}\right)\psi'(r)-\frac{1}{f}\left(B(\phi)-\frac{q^2 A_t^2}{f}\right)\psi(r)=-\frac{\omega^2}{f^2}\psi(r) \, .
\end{equation}
As in the zero temperature case, \eqref{mode} can be written in Schr\"odinger form,
\begin{equation}\label{eomschro}
-\frac{d^2}{d\xi^2}\tilde{\psi}+V_{Schr}\,\tilde{\psi}=\omega^2 \tilde{\psi} \, ,
\end{equation}
after a change of variable and field redefinition given by
\begin{equation}
\frac{d\xi}{dr}=\frac{1}{f}\, ,\quad \tilde{\psi}=\sqrt{h}\, \psi \, .
\end{equation}
The corresponding effective Schr\"odinger potential is then
\begin{equation}\label{schr3charge}
V_{Schr}=\frac{f^2 h''}{2h}-\frac{f^2 h'^2}{4 h^2}+\frac{ff'h'}{2h}+B(\phi) f-q^2 A_t \, .
\end{equation}
We emphasize once again that the background geometry will be unstable if the Schr\"odinger equation~\eqref{eomschro}
has a negative energy bound state $\omega^2<0$, corresponding to a solution which grows exponentially in time.
Furthermore, if there is an unstable mode $Im(\omega)>0$, then at the onset of the instability one should expect to find
a zero mode with $\omega = 0$.
Clearly, its profile will depend on the entire geometry, from the IR to the UV.

Indeed, the critical case is the zero energy state which corresponds to solutions of the equation
\begin{equation}\label{mode0}
\psi''(r)+\left(\frac{f'}{f}+\frac{h'}{h}\right)\psi'(r)-\frac{1}{f}\left(B(\phi)-\frac{q^2 A_t^2}{f}\right)\psi(r)=0 \, ,
\end{equation}
which therefore determines the zero modes.
After specifying the coupling $B(\phi)$, the critical temperature as a function of charge $q$
can be determined by solving~\eqref{mode0} numerically.
The two boundary conditions needed to fully specify the solution will be chosen as follows.
First, we will impose regularity at the horizon\footnote{Note that $\psi'(r_h)$ is fully determined by $\psi(r_h)$,
which  can be set to unity  due to the linearity of~\eqref{mode0}.}. The second boundary condition will come from specifying the UV asymptotics.
Indeed, as is well known, there are two modes in the UV $AdS_4$ region --
one is  interpreted as the source of the dual scalar operator, while the other as its expectation value.
Here we adopt the standard quantization, \emph{i.e.} choose the faster falloff to describe the expectation value, and the leading term to be the source.
Moreover, we will set the latter to zero, so that the $U(1)$ symmetry is broken spontaneously.
For a given temperature $T$, we expect such normalizable zero modes to appear at a special value of $q$.
Finally, we will work in the grand canonical ensemble by fixing the chemical potential -- in particular, we will  set it to $\mu=1$.

For concreteness in our numerics we choose the coupling to be
\begin{equation}\label{massfunB}
 B(\phi)=M^2 \cosh(\hat{\tau} \phi) \, ,
 \end{equation}
with $M$ and  $\hat{\tau}$  constant.
At the asymptotic boundary, where $r\rightarrow \infty$,  it behaves as
\begin{equation}
 B(\phi) \sim M^2 (1+\frac{\hat{\tau}^2}{2}\phi^2+\cdots)\, , \quad \text{as}\quad \phi\rightarrow 0 \, ,
 \end{equation}
and the leading terms in the UV expansion of $\psi$ are
\begin{equation}
\psi(r)=\frac{\psi^{(-)}}{r^{\Delta_-}}(1+\cdots)+\frac{\psi^{(+)}}{r^{\Delta_+}}(1+\cdots),\quad \Delta_{\pm}=\frac{3\pm\sqrt{3^2+4 M^2}}{2} \, .
\end{equation}
Since we are not allowing for a source term, we set $\psi^{(-)}=0$  in the expansion above.

On the other hand, in the extreme IR with $\phi\sim\ln(1/r)\rightarrow \infty$, $B(\phi)$ takes the form we have assumed in the previous sections,
\begin{equation}\
 B(\phi)=\frac{M^2}{2} e^{\hat{\tau}\phi}, \quad \text{as}\quad \phi\rightarrow \infty \, .
\end{equation}
Note that to obtain this relation we have assumed\footnote{Since the coupling~\eqref{massfunB}
is an even function of $\phi$, $\hat{\tau}<0$ gives nothing new.} $\hat{\tau}>0$.
It is helpful to point out that in the present case $\hat{\tau}$ is related to $\tau$ of~\eqref{scalB} by
\begin{equation}\
\hat{\tau}=-\frac{2}{\sqrt{3}}\tau \, .
\end{equation}
The special value $\hat{\tau}=0$ describes the case in which $B(\phi)=M^2$ is a constant.
Finally, the case in which the mass term $B(\phi) \, \psi^2$ scales in the same way as the kinetic term $(\p \psi)^2$
is obtained from~\eqref{scaletau} by choosing $m=3/4$,
\begin{equation}\label{scale}
\tau=2(m-1)=-\frac{1}{2}\quad \Rightarrow \quad\hat{\tau}=\frac{1}{\sqrt{3}} \, .
\end{equation}

One of the questions we are interested in is whether the superfluid instability in these models
can appear at arbitrary small values of the charge $q$.
Guided by our analytical estimates --  in terms of the effective mass in Section~\ref{masssection} or the Schr\"{o}dinger
potential in Section~\ref{SchrodingerSection} --  we will consider examples that address the following scenarios:
\begin{enumerate}
  \item Scaling case $\tau=2(m-1)$: We have a generalized BF bound, \eqref{generBF1} or \eqref{generBF2}, analogous to that of the standard $AdS_2$ case.
  If the bound is violated, the superfluid instability can be triggered for arbitrarily small charges,
  while if the bound is unbroken a minimal charge is required. The latter can be tuned by changing the location of the transition $\sim r_{tr}$ to
  the UV AdS geometry.
  \item Non-scaling case with $\tau<2(m-1)$ and $B_0<0$:
  As we discussed in case (i) of Section~\ref{nonscalneg}, the effective mass~\eqref{effM2} approaches negative infinity as $r\rightarrow 0$,
  and therefore the superfluid instability is expected to appear even at zero charge, \emph{i.e.} there is no $Q_{min}$.
  From the Schr\"{o}dinger potential~\eqref{schrpoten} standpoint, we see a large negative well
  as $r\rightarrow 0$. Therefore, the corresponding superfluid instability is expected to be associated with the far
  IR of the hyperscaling violating region.
  \item Remaining parameter ranges: A minimal charge is generically required in order to trigger a superfluid instability.
   For case (ii) of Section~\ref{nonscalneg},
   the effective mass~\eqref{effM2} approaches positive infinity as $r\rightarrow 0$.
  Similarly, the Schr\"{o}dinger potential~\eqref{schrpoten} is positive in the IR.
  The superfluid instability, if it is triggered, will be associated with the effective UV geometry of the hyperscaling violating regime, and not
  with its IR regime.
  In case (iii) of Section~\ref{nonscalneg}, the potential~\eqref{schrpoten} becomes and stays positive as one gets sufficiently close to
  $r = 0$, no matter how large the values
  of $B_0$ and $Q$ are -- even when $B_0<0$. Unlike the case we have just discussed, however, the potential remains finite and instabilities
  can still be triggered, but are associated with the $r\sim r_{tr}$ transition region. In these cases a minimal charge is needed to ensure that
  $V_{Schr}$ has a sufficiently negative region.
\end{enumerate}
Below we will provide concrete examples realizing each of these scenarios.
In particular, we will investigate~\eqref{mode0} numerically in order to determine
the critical temperature associated with the zero mode solutions as a function of the charge $q$.

\subsection{Scaling case}
The scaling case corresponds to $\hat{\tau}=1/\sqrt{3}$, so that our mass coupling is given by
\begin{equation}
 B(\phi)=M^2 \cosh(\phi/\sqrt{3}) \, .
 \end{equation}
The equation of motion for $\psi$ on the background geometry~\eqref{ir3charge} then becomes
\begin{equation}
\psi''(\rho)-\frac{1}{\rho}\psi'(\rho)-\frac{2(3 M^2\rho^2-2q^2)}{9\rho^4}\, \psi(\rho)=0 \, ,
\end{equation}
and is solved by\footnote{We point out that~\eqref{scalepsi} holds
when $\nu$ is not an integer. However, when $M^2=3 (k^2-1)/2$ with $k$ an integer, the solution is given by
\begin{equation}
\psi(\rho)=\frac{\rho}{q}\left[c_1  J_{k}\left(\frac{2q}{3\rho}\right)+c_2  Y_{k}\left(\frac{2q}{3\rho}\right)\right].
\end{equation}}
\begin{equation}\label{scalepsi}
\psi(\rho)=\frac{\rho}{q}\left[c_1\, \Gamma (1-\nu) J_{-\nu}\left(\frac{2q}{3\rho}\right)+c_2\, \Gamma (1+\nu)
J_{\nu}\left(\frac{2q}{3\rho}\right)\right]\, ,
\end{equation}
where the index $\nu=\sqrt{1+2 M^2/3}$.
The instability associated with the index becoming imaginary, when $M^2<-3/2$, is
equivalent to the violation of the BF bound \eqref{generBF1}, with the parameter choice $m=3/4$ and $n=1/4$.
Notice that in this case $\nu$ does not contain any charge dependence, unlike the standard $AdS_2$ case~\eqref{ads2BF}.

\begin{figure}
\begin{center}
\includegraphics[width=3.5in]{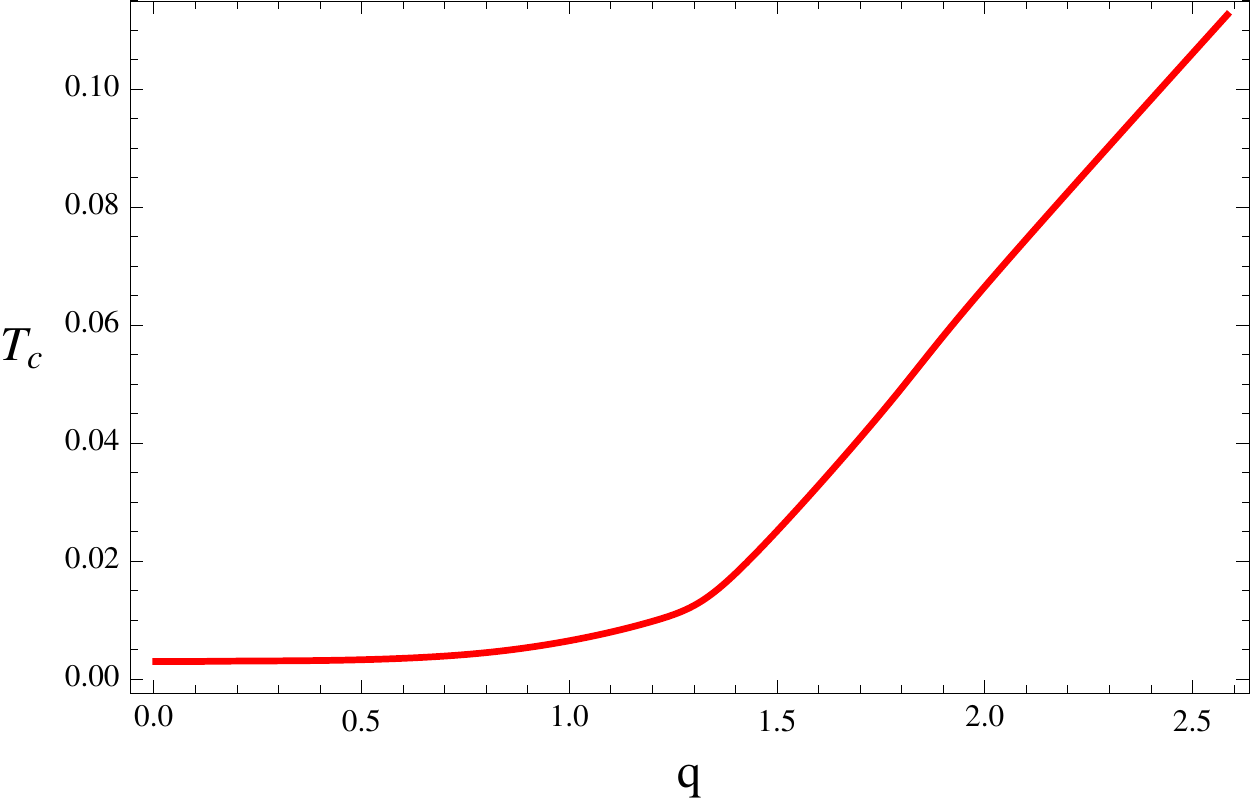}
\caption{Critical temperature as a function of charge $q$ for the scaling case with $M^2=-2$ and $\hat{\tau}=1/\sqrt{3}$.
The critical temperature at $q=0$ is $T_c\approx 0.00297$. There is no minimal charge, thus a neutral scalar will condense.
We work in units in which the chemical potential is $\mu=1$.}
\label{fgscale}
\end{center}
\end{figure}

We will consider two qualitatively different cases,
by choosing first $M^2=-2$, for which $\nu=\sqrt{-1/3}$ is complex, and then $M^2=-5/4$, i.e. $\nu=\sqrt{1/6}$ real.
The critical temperature of the zero mode solutions as a function of charge $q$ for $M^2=-2$ is presented in figure~\ref{fgscale}.
As one can see, $T_c$ decreases as we lower the charge $q$, but the zero mode survives even when the charge is zero. In that case the
 instability is due to the breaking of the local BF bound in the far IR of the hyperscaling violating geometry, where the charge
term is negligible.
Thus, here we see a model which gives rise to a superfluid condensate at arbitrarily
small values of the charge, in accordance with the analogous $AdS_2$ result.

\begin{figure}
\begin{center}
\includegraphics[width=3.5in]{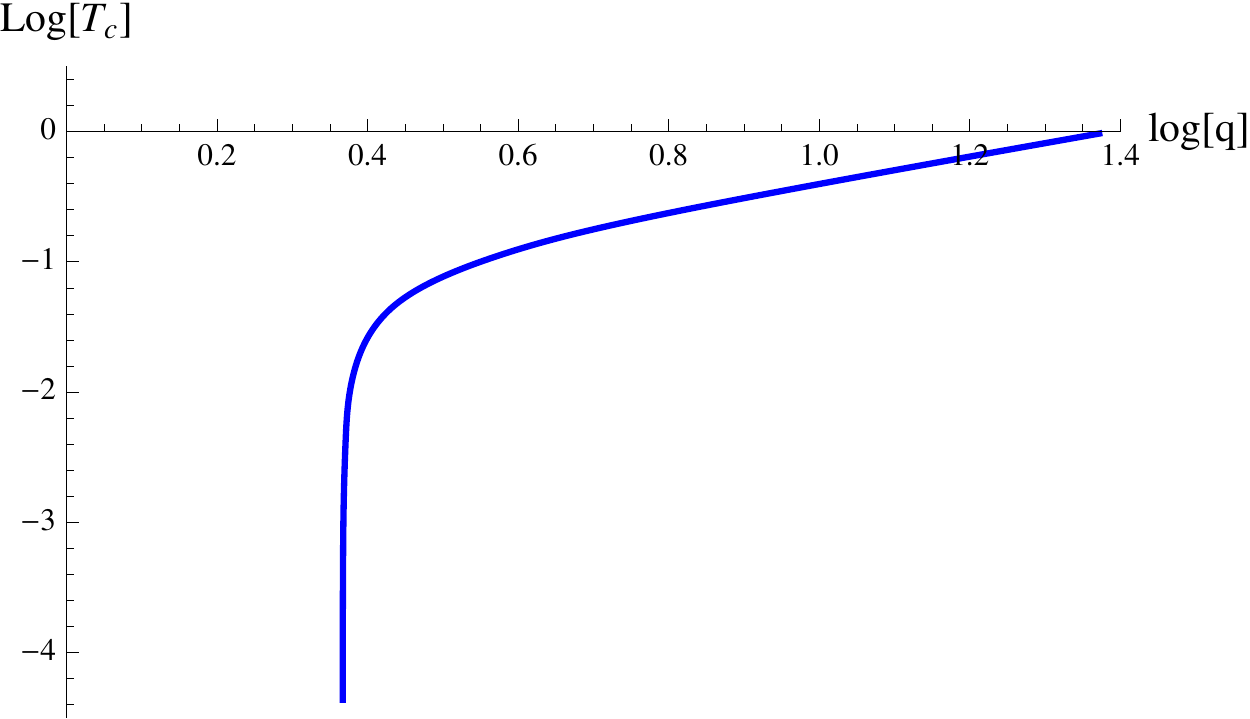}
\caption{Critical temperature versus charge $q$ for the scaling case with $M^2=-5/4$ and $\hat{\tau}=1/\sqrt{3}$.
We see a minimal charge $q\approx 2.327$ below which the zero mode for superfluid instability does not exist.
We work in units in which the chemical potential is $\mu=1$.}
\label{fgscalemin}
\end{center}
\end{figure}

In figure~\ref{fgscalemin} we show the critical temperature as a function of $q$ for $M^2=-5/4$.
In this case the index $\nu$ is real and the corresponding BF bound is unbroken.
Just as expected, there is a minimal charge at which the background
will become unstable to developing non-trivial scalar hair.
We note that although the existence of a minimal charge is analogous to what would occur in AdS, the behavior of the
charge term in the hyperscaling violating geometries is not -- it is most important near $r_{tr}$ and negligible in the far IR.

\subsection{Non-scaling case with infinitely negative effective mass}
Here we are considering the scenario discussed in case (i) of Section \ref{nonscalneg}.
The effective mass~\eqref{effM2} approaches negative infinity as $r \rightarrow 0$.
Thus, the expectation is that the zero mode should survive at arbitrarily small values of the charge.
We consider the following coupling
\begin{equation}
 B(\phi)=-2 \cosh(\sqrt{3}\phi),
 \end{equation}
which is obtained from~\eqref{massfunB} by choosing $M^2=-2$ and $\hat{\tau}=\sqrt{3}$.

\begin{figure}
\begin{center}
\includegraphics[width=3.5in]{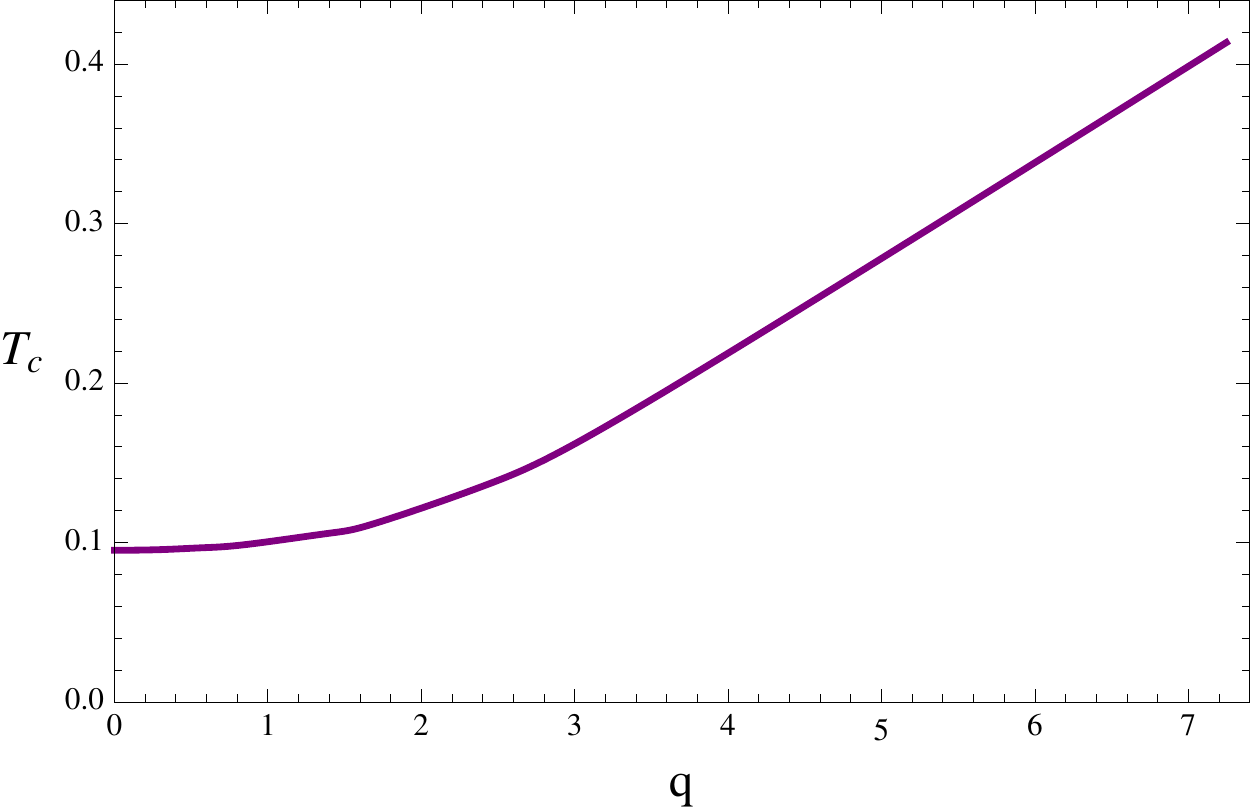}
\caption{Critical temperature as a function of charge $q$ for the non-scaling case with $M^2=-2$ and $\hat{\tau}=\sqrt{3}$.
The critical temperature at $q=0$ is $T_c\approx 0.0951$. We work in units with $\mu=1$.}
\label{fgnonscaleneg}
\end{center}
\end{figure}

Figure~\ref{fgnonscaleneg} shows the critical temperature as a function of charge $q$. One can clearly see that there is a phase transition
even in the limit of zero charge.
It is helpful to compare this case to the scaling one with $M^2=-2$, as they both share the same UV mass.
Since the effective mass in the far IR goes to negative infinity, in the present case
(with $\hat{\tau}=\sqrt{3}$) instabilities should be triggered much more easily than in the scaling one (with $\hat{\tau}=1/\sqrt{3}$).
As a result, we expect the critical temperature here to be higher than that of the scaling scenario.
This is precisely what we find from the numerics by comparing figure~\ref{fgnonscaleneg} with figure~\ref{fgscale}.
Notice that this behavior is independent of how large $B_0 \sim M^2$ is, a feature due entirely to the non-trivial coupling $B(\phi)$, which
is absent in the standard holographic superconductor scenario.
Thus, by appropriately choosing the functional dependence of the coupling, we can facilitate the phase transition and increase $T_c$.

\subsection{Remaining cases}
In the remaining cases we discussed in Section \ref{nonscalneg}, we don't expect the zero mode to
exist at arbitrarily small values of $q$. Let's focus on the choice
\begin{equation}\label{nonmass}
B(\phi)=M^2\, ,
\end{equation}
which corresponds to $\hat{\tau}=0$ (or equivalently $\tau=0$) and falls under the category (iii) of Section \ref{nonscalneg}.
Before discussing the numerics, we stress that
in these cases it's hard to identify sharp analytical instability criteria in the scaling regime.
The equation for the $\psi$ perturbation now reads
\begin{equation}
\psi''(\rho)-\frac{1}{\rho}\psi'(\rho)-\frac{4(\sqrt{3} M^2\rho-q^2)}{9\rho^4} \, \psi(\rho)=0,
\end{equation}
and the solution in general is given by
\begin{equation}
\psi(\rho)=e^{\frac{2 i q}{3\rho}}\left[c_a\, {}_1 F_1\left(\frac{3}{2}-\frac{i\, M^2}{\sqrt{3}q},3,\frac{4 i q}{3\rho}\right) +
c_b\, U\left(\frac{3}{2}-\frac{i\, M^2}{\sqrt{3}q},3,\frac{4 i q}{3\rho}\right)\right],
\end{equation}
where ${}_1F_1(a,b,x)$ is the Kummer confluent hypergeometric function and $U(a,b,x)$ is a confluent hypergeometric function.
The special case with $q=0$ needs to be treated separately, and is
\begin{equation}
\psi(\rho)=\rho \left[c_a\,  I_{2}\left(\frac{4}{3^{3/4}}\sqrt{\frac{M^2}{\rho}}\right)+
c_b\,  K_{2}\left(\frac{4}{3^{3/4}}\sqrt{\frac{M^2}{\rho}}\right)\right]\, .
\end{equation}
In contrast to the scaling case~\eqref{scalepsi}, it is not immediately apparent how to extract information about potential instabilities
from the structure of the solutions. In particular, there is no simple analog of the generalized BF bound, illustrating
the challenge of obtaining generic analytical conditions
for the onset of the phase transition.

\begin{figure}
\begin{center}
\includegraphics[width=3.5in]{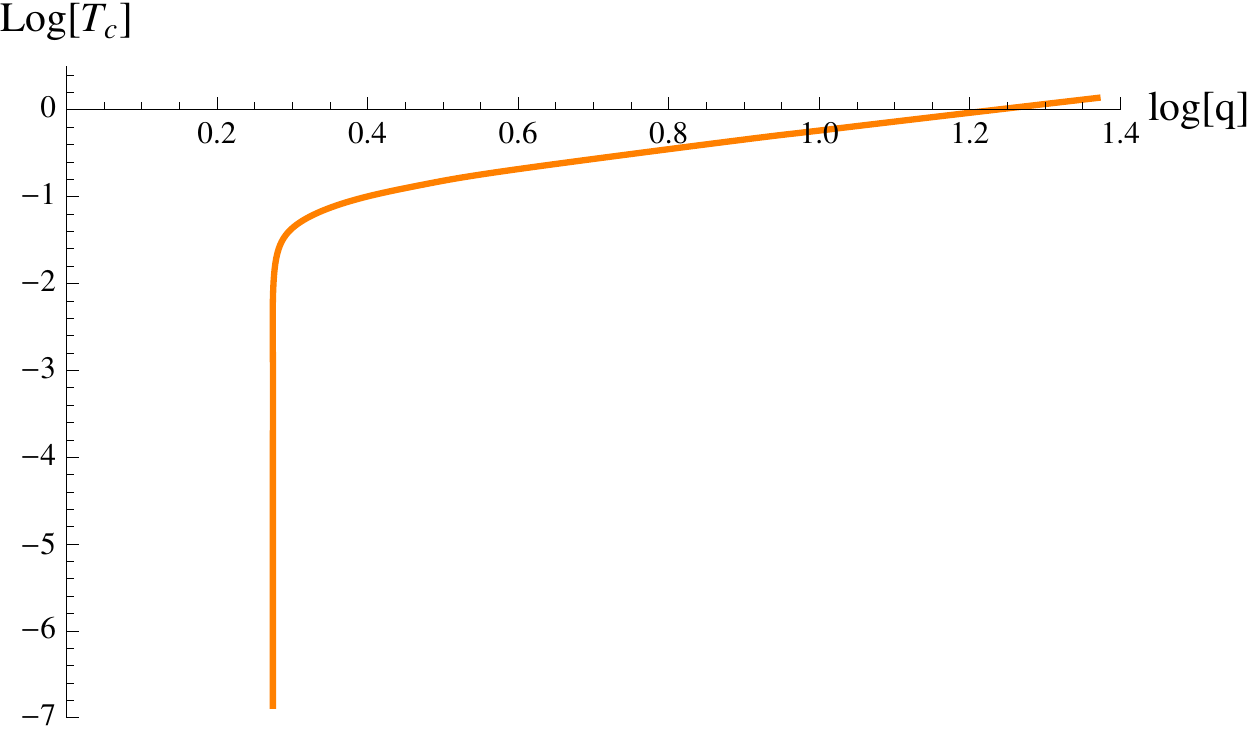}
\caption{Log-Log plot of the critical temperature versus charge $q$ for the non-scaling case with $M^2=-2$ and $\hat{\tau}=0$.
The critical temperature goes to zero at $q\approx1.88$. We work in units with $\mu=1$.}
\label{fgnoscale}
\end{center}
\end{figure}

\begin{figure}
\begin{center}
\includegraphics[width=4.8in]{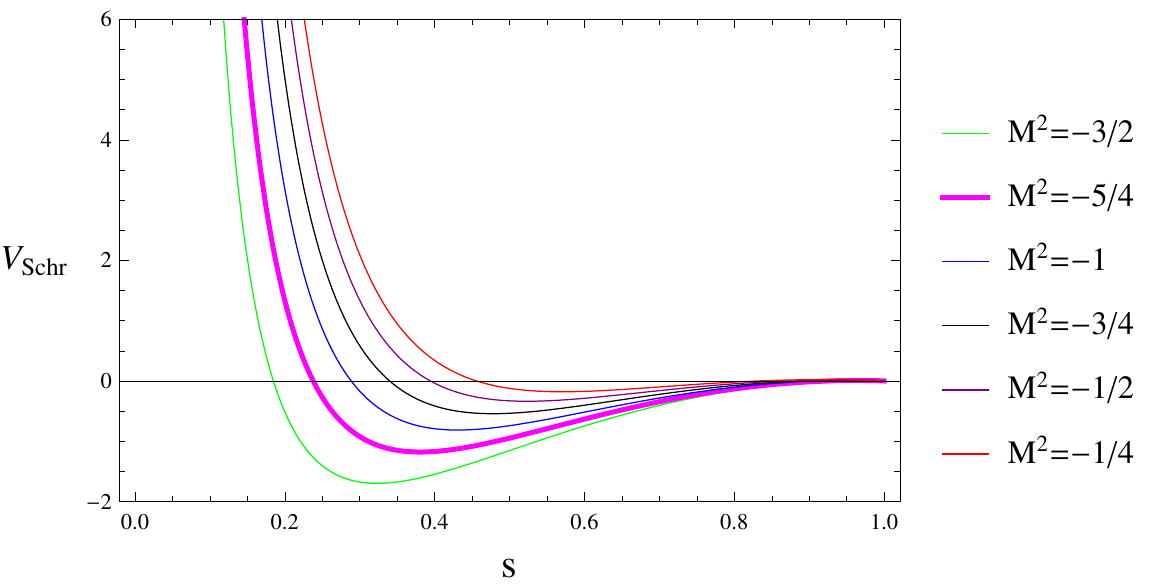}
\caption{Schr\"{o}dinger potential~\eqref{schr3charge} as a function of radial coordinate $s=\frac{r_h+Q}{r+Q}$ for the
non-scaling case $\hat{\tau}=0$. The different curves have the same values of charge $q\approx 2.6002$ and temperature $T\approx 1.378\times 10^{-4}$
but different values of $M^2$.
The horizon is located at $s=1$ and the UV $AdS_4$ boundary at $s=0$.
We choose parameters such that the thick magenta line corresponds to the zero mode solution for this particular choice of $q$ and $T$.
We work in units of $\mu=1$.}
\label{fgschr}
\end{center}
\end{figure}

The behavior of the critical temperature as a function of charge $q$ for the choice $M^2=-2$ is presented in figure~\ref{fgnoscale}.
Although we can not solve the system at very low temperatures, we find strong evidence that a minimal
charge is indeed required in order to trigger the superfluid instability, as indicated by the effective mass and Schrodinger potential analysis.
\begin{figure}
\begin{center}
\includegraphics[width=3.5in]{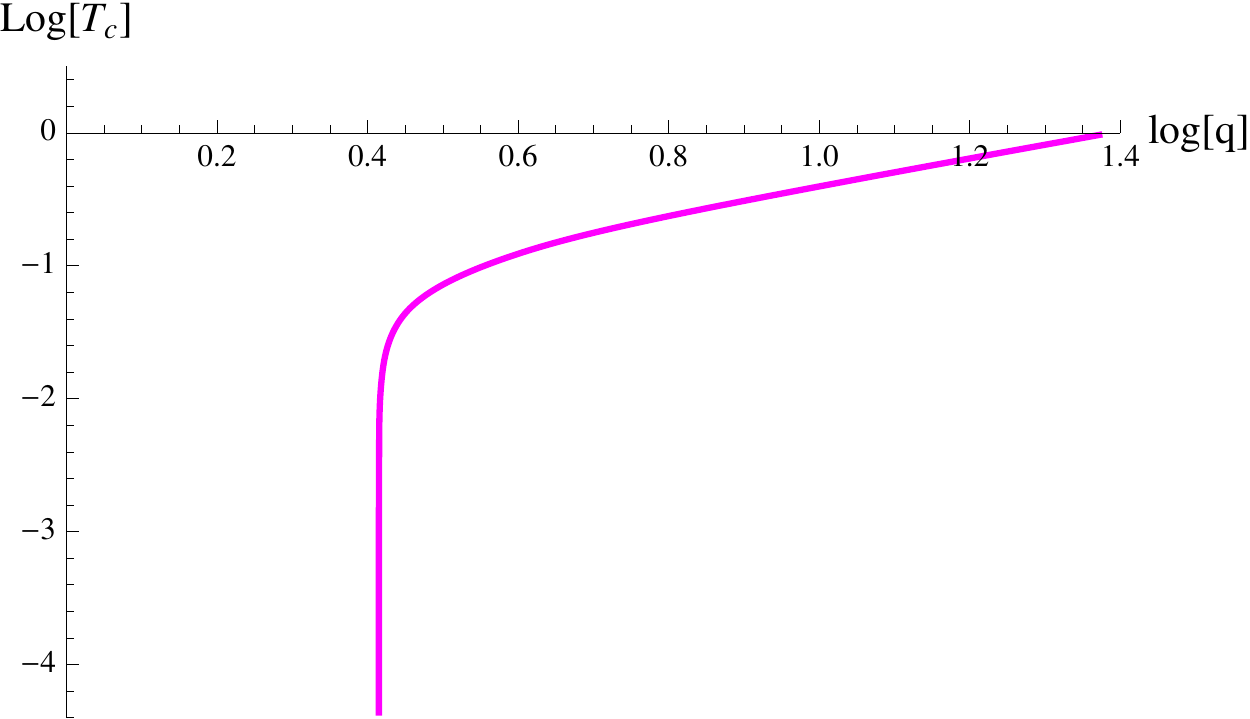}
\caption{Log-Log plot of the critical temperature as a function of charge $q$ for the non-scaling case with $M^2=-5/4$ and $\hat{\tau}=0$.
The critical temperature goes to zero at $q\approx 2.6$. We work in units with $\mu=1$.}
\label{fgnoscalebig}
\end{center}
\end{figure}
A bigger value of $M^2=-5/4$ is then shown in figure~\ref{fgnoscalebig}.
The qualitative behavior is very similar to that of figure~\ref{fgnoscale}.
However, we note that as we increase the size of $M^2$,
a bigger minimal charge is required in order to trigger the superfluid instability.
This point can be understood qualitatively by comparing the Schr\"{o}dinger potential~\eqref{schr3charge}
for different values of the mass~\eqref{nonmass} but keeping $q$ and $T$ fixed, as is done in figure \ref{fgschr}.
Recall that we are working in the grand canonical ensemble and have therefore fixed the chemical potential to $\mu=1$.

We choose parameters such that the thick magenta line in figure~\ref{fgschr} corresponds to the zero mode
solution of~\eqref{mode0} for $M^2=-5/4, \hat{\tau}=0$ at $q\approx 2.6002$ and $T\approx 1.378\times 10^{-4}$.
From~\eqref{wkb}, this case gives the smallest negative potential well which supports a zero mode bound
state for the chosen values of $q$ and $T$ (e.g. for $k=1$ of~\eqref{wkb}).
One can in principle change $M^2$ to obtain a much larger negative potential region such that~\eqref{wkb} can be 
satisfied for $k\geqslant 2$.
However, it is easy to see from figure~\ref{fgschr} that the range in which
the Schr\"{o}dinger potential is  negative as well as its depth becomes smaller and smaller as one increases $M^2$.
Therefore, in order to support a zero mode, a bigger value of $q$ is required to compensate for the increase in the mass parameter $M^2$.
In addition, we note that there is a positive potential region in the deep IR, which is too small to see from figure~\ref{fgschr}.

\begin{figure}[h!]
\begin{center}
\includegraphics[width=3.5in]{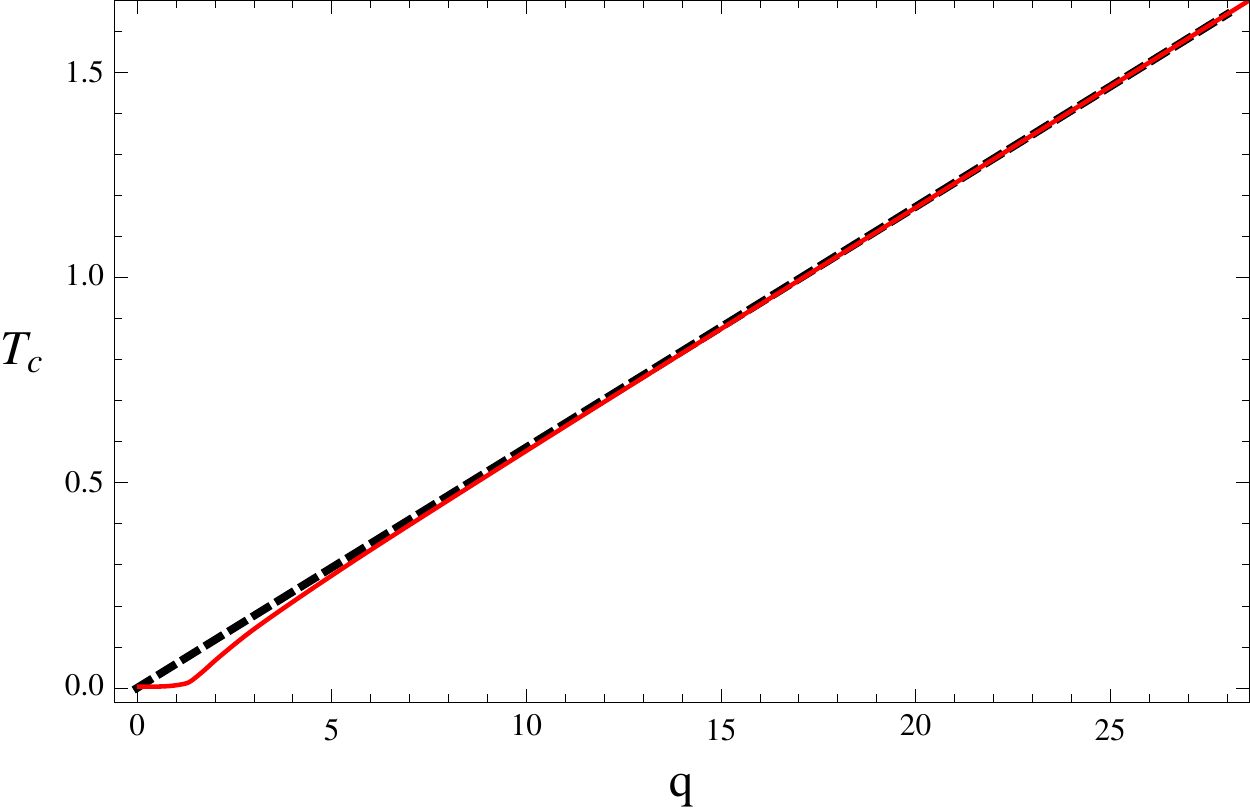}
\caption{The red solid line is the critical temperature as a function of charge $q$ for the scaling case with $M^2=-2$ and
$\hat{\tau}=1/\sqrt{3}$. The dashed black line corresponds to the probe limit result with $T_c/q\approx 0.586$. We work in units of $\mu=1$.}
\label{fgprob}
\end{center}
\end{figure}

Before closing this Section, we would like to point out one final feature visible from the numerics.
As one can see from inspecting figures~\ref{fgscale} to~\ref{fgnoscalebig},
when $q$ is large the value of $T_c$ increases linearly with $q$.
This behavior can be understood as follows~\cite{Cai:2013aca}.
Taking $q\rightarrow \infty$ while keeping $q\Psi$ and $q A_\mu$ finite, we arrive at the probe
limit in which the gauge field and the charged scalar do not backreact on the background geometry.
In order to compare our results with those in the probe limit, we have to perform
the scaling transformation $\Psi\rightarrow q\Psi$
and $A_\mu\rightarrow q A_\mu$.
After taking these rescalings into account, the physical dimensionless temperature becomes
$T_c/q\mu$. Since we are working with $\mu=1$ (recall that we are in the grand canonical ensemble),
this tells us that $T_c \propto q$, which is precisely what is observed from the numerics
in the large charge limit.
The backreacton of the $U(1)$ field and charged scalar on the geometry becomes smaller and smaller as $q$ is increased, explaining
again why we observe a linear behavior for $T_c$ when the charge is large. We confirm this in figure~\ref{fgprob}, which
has the same choice of parameters as figure~\ref{fgscale}, but reaches higher values of $q$.
It is clear that the large $q$ behavior can be well approximated by the linear function $T_c=\gamma q$ with $\gamma$ a
constant, as expected from the probe limit argument. For small charges we deviate from the linear relationship, as clearly visible from
figure~\ref{fgscale} as well as figure~\ref{fgprob}.


\begin{acknowledgments}
We are grateful to Blaise Gout$\acute{e}$raux, Elias Kiritsis and Jim Liu for many insightful comments on the draft.
We would also like to thank Tom Hartman for useful conversations.
The work of LL was supported in part by European Union's Seventh Framework Programme under grant agreements (FP7-REGPOT-2012-2013-1) no 316165 and the Advanced ERC grant SM-grav 669288.

\end{acknowledgments}

\end{document}